\documentclass[final,1p,times]{elsarticle}
\usepackage{amsmath,amssymb,amsfonts}
\usepackage{algorithm}
\usepackage[noend]{algpseudocode}
\usepackage{url}
\usepackage{tabto}
\usepackage{graphicx}
\usepackage{textcomp}
\usepackage{xcolor}
\usepackage{subcaption}

\usepackage{multirow}

\journal{XXX}

\begin{document}
\newtheorem{theorem}{Theorem}[section]
\newtheorem{lemma}[theorem]{Lemma}
\newtheorem{definition}[theorem]{Definition}

\begin{frontmatter}

\title{A Comprehensive Scalable Framework for Cloud-Native Pattern Detection with Enhanced Expressiveness}

\affiliation[auth]{organization={Aristotle University of Thessaloniki, Dept. of Informatics},
            city={Thessaloniki},
            country={Greece}}
\author[auth]{Ioannis Mavroudopoulos}
\ead{mavroudo@csd.auth.gr}
\author[auth]{Anastasios Gounaris}
\ead{gounaria@csd.auth.gr}

\begin{abstract}
Detecting complex patterns in large volumes of event
logs has diverse applications in various domains, such as business
processes and fraud detection. Existing systems like ELK are
commonly used to tackle this challenge, but their performance
deteriorates for large patterns, while they suffer from limitations
in terms of expressiveness and explanatory capabilities for their
responses. In this work, we propose a solution that integrates a
Complex Event Processing (CEP) engine into a broader query
processsor on top of a decoupled storage infrastructure containing
inverted indices of log events. The results demonstrate that
our system excels in scalability and robustness, particularly in
handling complex queries. Notably, our proposed system delivers
responses for large complex patterns within seconds, while ELK
experiences timeouts after 10 minutes.
 It also significantly outperforms solutions relying on FlinkCEP and executing MATCH\_RECOGNIZE SQL queries.
\end{abstract}

\end{frontmatter}

\section{Introduction}
In recent years, logging systems have become an essential tool for organizations to monitor their processes, generating log entries that can reach the order of terabytes per day\cite{seattle_report}. Each log entry typically contains at least an event type and a timestamp, usually along with other domain-specific information, thus the set of logs forms an event sequence database. The advanced analytics that are then applied to these sequences typically fall into two main categories: sequential pattern mining (SPM) \cite{fournier2017survey} and complex event processing (CEP) \cite{sase+,sase_2014}. SPM techniques extract interesting sequences from the event sequence database, such as frequent\cite{Zaki01}, important, and highly utility ones \cite{fournier2019survey}, while CEP detects arbitrary patterns in real-time over a stream of events.

However, organizations often need to detect specific behaviors that are not known beforehand or are not frequent (arbitrary patterns); there is a need to support ad-hoc pattern queries over these very large sequence databases. For instance, an online shop may want to identify customers who purchase product A followed by product B and are likely to be interested in the newly arrived product C, in order to efficiently make a targeted advertisement. Similarly, in case of a cybersecurity attack, companies need to efficiently search their logs for event sequences to determine whether they have been affected by the attack.

This problem can be addressed by employing a CEP engine, such as SASE~\cite{sase_2014}, and treating the event sequence database as an event stream. However, findings from \cite{mavroudopoulos2021sequence} strongly indicate that this approach is suboptimal when applied to large sequence databases. The reason behind this suboptimal performance is that CEP engines require parsing the entire dataset for each query, as they lack any indexing mechanisms to facilitate efficient search space reduction. An alternative solution involves the utilization of MATCH\_RECOGNIZE (MR) \cite{match_rec_trino,match_rec_snowflake}, a recent SQL extension that is capable of pattern detection, similar to CEP. Nevertheless, current implementations of MR do not rely on any tailored indexing either, and thus they encounter similar limitations as CEP, i.e. they cannot scale in big data scenarios. Therefore, a form of indexing is crucial in order to effectively address the detection of arbitrary patterns in large volumes of data.

The main problem this work aims to address is detecting arbitrary complex patterns in large log databases. In the online shopping example described above, the pattern is a simple one and can be denoted as $<A,B,C>$. Complex patterns include negations and iterations, e.g., $ < A, B, !C, D, E*, A, B+, C, D, E > $, while duration constraints between events may apply, e.g., the timestamps between the first and the last event should not exceed a threshold. We propose a novel approach, according to which the complex pattern above in our experimental setting can be detected in less than 10 seconds, whereas the open-source state-of-the-art ELK (Elasticsearch, Logstash, Kibana) stack\footnote{\url{https://www.elastic.co/what-is/elk-stack}} times out after 10 minutes. Note that these complex patterns are already supported by both CEP engines, like SASE~\cite{sase_2014} and FlinkCEP~\cite{flinkcep}, and MR; however, as also evidenced by our experimental results, these solutions do not scale.
The second problem we target is how to offer explainability in simple patterns, i.e., to explain to the user the minimal change in the timestamps required for a trace to include the requested pattern in case the trace contains the pattern events but not in the user-defined order; this increases the trust in the system \cite{whyNot}.

Both problems are addressed through the development of a system that, to the best of our knowledge, is the first one that encapsulates a CEP engine within a query processor that is totally decoupled from the (indexed) log storage thus adhering to the modern design principles for cloud-native big data management systems \cite{TGP+19}.
As a basis, we use the indexing solution of the preliminary version of the system \cite{siesta}, which is shown to outperform other index-based approaches, such as \cite{nanopoulos_2002}, \cite{lcjoin} and ELK, for simple patterns. 
Building on top of \cite{siesta} indexing rationale, we fully re-engineer its query processor to (i) encapsulate SASE \cite{sase+, sase_2014}, (ii) run in Spark to attain scalability and (iii) offer explanations. We also heavily modify the storage layer departing from Cassandra (so that storage and processing are fully decoupled) while addressing incremental indexing issues.  In summary, this work has four main contributions. 
\begin{itemize}
	\item We propose a novel architecture for arbitrary pattern detection in log databases that encapsulates a CEP engine and its storage is decoupled from the query processing module.
	\item We propose a methodology to efficiently process complex queries and offer explanation when matches are not found. 
	\item We conduct experiments using a large volume of data to demonstrate the superior scalability and pattern detection capabilities of SIESTA compared to its competitors, including ELK, FlinkCEP, and MATCH\_RECOGNIZE.
	\item We provide the complete solution in open source and the experimental setting for reproducibility of our work. 
\end{itemize}

The remaining sections of this paper are organized as follows. We present the related work next. In Section~\ref{sec:background}, we introduce the notation and briefly describe some of the key components of SIESTA. Section~\ref{sec:AQP} discusses the novel aspects of the query processor. In Section~\ref{sec:storage}, we present an alternative storage solution to the one in the preliminary version of SIESTA, the incremental indexing approach, and additional storage optimizations.
In Section~\ref{sec:evaluation}, we evaluate the proposed implementation. Section~\ref{sec:future-work} presents some ideas for future extensions, and we conclude the paper in Section~\ref{sec:conclusion}.

\section{Related Work}
\label{sec:related work}
SIESTA relates more closely with three other areas, which are briefly described in turn. 

\emph{Pattern Detection and expressiveness of CEP systems:}
There have been numerous surveys presenting scalable solutions for detecting complex events in data streams, with a particular focus on pattern detection ~\cite{cugola2012,dayarathna2018}. Over the years, several general-purpose CEP languages have been proposed, offering different capabilities and levels of expressiveness in terms of supported pattern queries \cite{giatrakos2020vldb}. 
Preliminary languages, like SASE~\cite{sase_2006} and SQL-TSQ\cite{sadri_2001}, could only support a single selection policy, namely strict-contiquity (SC). Later, an extension called SASE+~\cite{sase+} was introduced, which allowed the skipping of irrelevant events in the stream, enabling skip-till-next-match (STNM) and skip-till-any-match (STAM) policies. Moreover, SASE+ expanded upon the original version by supporting the Kleene* operator. 
Based on the SASE+ language, the CEP engine SASE has been developed~\cite{sase_2014}. We chose to integrate it with SIESTA as it is lightweight software and provides support for a wide variety of patterns. To the best of our knowledge, such usage of a CEP engine has not appeared before, while we avoid reinventing the wheel.
Other examples of CEP engines include Cayuga\cite{demers2007cayuga}, NextCEP\cite{schultz2009distributed}, and FlinkCEP\cite{flinkcep}, which is a complex event recognition and processing on top of Apache Flink\cite{flink15}. 
In addition to traditional pattern detection, CEP systems have also tackled real-world challenges such as handling imprecise event timestamps~\cite{zhang_2013_impreccise} and out-of-order event arrivals~\cite{carbone2020beyond}. 

The MATCH\_RECOGNIZE (MR) SQL extension can express complex queries, similar to CEP engines, but it operates over static data, i.e., data stored in a relational database. MR is supported by various query engines, including Trino \cite{match_rec_trino} and Snowflake \cite{match_rec_snowflake}. We directly compare our solutions against such an implementation. 
Notably,  recently, a system called TReX~\cite{huang2023t} that builds upon MR and optimizes pattern searches in time series data leveraging query plan optimization techniques has been developed. These advances refer to execution plans, where event types are derived from time series data on the fly and the corresponding low-level operators can be re-ordered based on metadata, such as selectivity. Consequently, they are orthogonal to our work, which assumes that event type instances are already present in the logs. However, in our solution, we also extract occurrences of pairs of event types ordered by their selectivity.

\emph{Pattern mining:} For non-streaming data, a series of methods have been developed in order to mine patterns. The majority of these proposals look for frequent patterns; e.g.,  Sahli et al in \cite{acme_2014} proposed a scalable method for detecting frequent patterns in long sequences. As another example, in several other fields such as biology, several methods have developed, which are typically based on statistics (e.g., \cite{madmax_2009,varun_2011}) and suffix trees (e.g.,  \cite{flame_2010}). Parallel flavors have also been proposed, e.g., in \cite{psmile_2004,WU202131}, with the latter one implemented on top of Apache Hadoop. Other forms of mined patterns include outlying patterns \cite{top_2019}, high contrast patterns \cite{9460823} or general patterns with high utility as shown in \cite{uspan_2012}. \cite{aoga2017mining} support efficient pattern detection of frequent patterns by building indices. Finally, 
extensive research has been conducted to detect patterns under various constraints~\cite{pei2007constraint}. Introducing constraints not only benefits users in finding precisely what they need but also enhances the efficiency of the methods by avoiding unnecessary computations. Inspired by the advances in these relative fields, our aim in this work is to enable the efficient detection of more complex queries over large volumes of data.

\emph{Business processes:} There are applications in business process management that employ pattern mining techniques to find outlier patterns and clear log files of infrequent behavior, e.g., \cite{detection_and_removal_2020, huang_ying_2018}, in order to facilitate process discovery. Another application is to predict the upcoming events in a trace \cite{9325056,prediction_2012,FrancescomarinoG22} and also use this information to determine if a trace will fail to execute properly \cite{failure_prediction_2019}. All these techniques complement the proposed framework, and it would be interesting, in the future, to evaluate the performance of a domain-specific engine tailored to this context, rather than using SASE or any other competitor against which we already compare. Such an engine may offer more specialized capabilities for business process management.
Finally, in the field of Business Process Mining, the MR operator has been employed to detect patterns in log files \cite{augusto2021efficient}. 



\section{Background}
\label{sec:background}
We begin with a brief description of the notation and an overview of the SIESTA system.
SIESTA is an infrastructure for efficient support of sequential pattern queries based on inverted indices, generated from a logfile $L$ containing timestamped events $E$. The events are of a specific type and are logically grouped into sets called cases, sessions, or traces\footnote{We use the terms trace, case and session interchangeably.}. More formally:

 \begin{definition}{\textbf{(Event Log)}}
 	\label{def_eventLog}
 	Let $A$ be a finite set of activities (aka tasks). A log $L$ is defined as $L=(E,C,\gamma,\delta,\prec)$ where $E$ is the finite set of events, $C$ is the finite set of Cases, $\gamma:E \rightarrow C $ is a surjective function assigning events to Cases, $\delta:E\rightarrow A $ is a surjective function assigning events to activities. An event $ev \in E$ that belongs in a case $t\in C$, is a tuple that consists of at least a recorded timestamp $ts$, denoting the recording of task execution, an event type $type \in A$,  and a position $pos$, denoting the position of the event in $t$ (i.e., $t = <ev_1, ev_2,\dots, ev_n>$ and $ev_i.pos = i$ ). Finally, $\prec$ is the strict total ordering over events belonging to a specific case, deriving from the events' timestamp.
 
\end{definition}

Within a specific case, we assume that two events cannot have the same timestamp. In sessions like web user activity and similar ones, events are usually instantaneous. However, this is not the case in task executions in business processes. In the latter case, the timestamp refers to either the beginning of the activity or its completion, but in any case, logging needs to be consistent. 
Table \ref{table:symbol_int} summarizes the main notation.

\begin{table}[tb!]	
	\centering
		\caption{Frequently used symbols and interpretation}
		\begin{tabular}{|p{0.15\columnwidth}|p{0.7\columnwidth}|}
			\hline
			{\bf Symbol} & {\bf Short description} \\ \hline	
			$L$ & the log database containing events \\ \hline
			$A$ &  the set of activities (or tasks), i.e., the event types \\ \hline
			$E$  &  the set of all event instances \\ \hline
			$C$ & the set of cases, where each case corresponds to a single logical unit of execution, i.e., a session, a trace of a specific business process instance execution, and so on \\ \hline
			$ev$ & an event ($ev \in E$), which is an instance of an event type \\ \hline
			$ts$ & the timestamp of an event (also denoted as $ev.ts$) \\ \hline
			$pos$ & the position of an event in its trace (also denoted as $ev.pos$)\\ \hline	
			$type $ & the event type of an event (also denoted as $ev.type$); $type\in A$\\ \hline 
			$s$ & CEP operator, $s \in [\_,+,*,!,|| ]$\\ \hline
			$e$ & query event in QP, $e = (a,s)$, $a\in A$\\ \hline
		\end{tabular}
	\label{table:symbol_int}
\end{table}

\begin{definition}[\textit{et-pair}]
    \label{def:et-pair}
	An event type pair, or \textit{et-pair} for short , is a pair $(a_i,a_j)$, where $a_i,a_j \in A$
\end{definition}

\begin{definition}[\textit{event-pair}]
	\label{def:event_pair}
	For a given \textit{et-pair} $(a_i,a_j)$ and a sequence of events $t = <ev_1,ev_2,...,ev_n>$, there is an \textit{event-pair} $(ev_x,ev_y)$ with  $x\le y \land ev_x.type = a_i \land ev_y.type = a_j$ and $ev_x,ev_y \in t$. Note that based on this definition, the events in a \textit{event-pair} do not have to be consecutive.
\end{definition}

\begin{definition}[\textit{event-pairs} non-overlapping in time]
	\label{def:non_overlap}
	Two \textit{event-pairs} $(ev_x,ev_y)$ and ($ev_i,ev_j$) are non-overlapping in time if and only if $x > j \lor y < i$. 
\end{definition}

In the original SIESTA, event-pairs are generated based on Definition~\ref{def:non_overlap}. However, in our proposed solution, we allow for two \textit{event-pairs} to partially overlap in time. This modification affects only the \textit{event-pairs} ($ev_1,ev_2$) where $ev_1.type=ev_2.type=a_i$ (i.e., both event types are the same), in which case the second event can be reused as the first event in another \textit{event-pair} instance.  This enables us to support efficiently more advanced queries, like time constraints, which will be discussed in Section~\ref{sec:AQP}.

\begin{figure}[tb!]
	\centering
	\includegraphics[width=0.9\columnwidth]{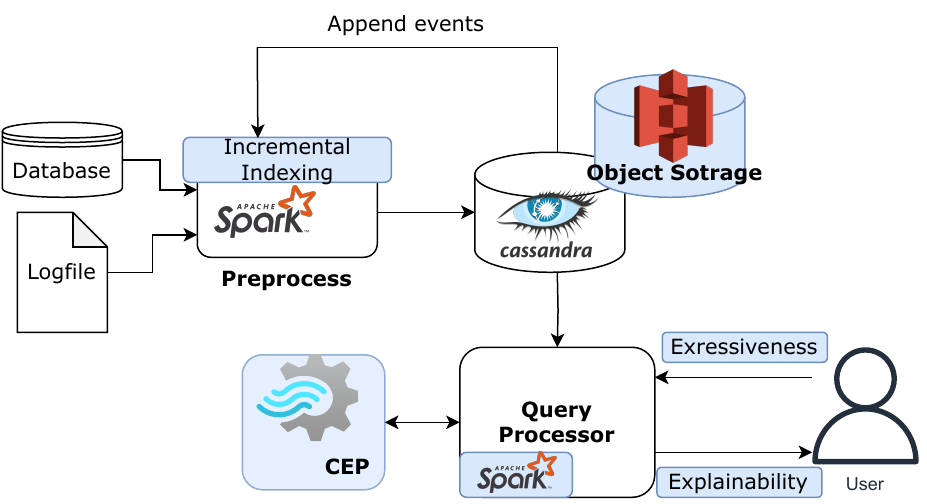}
	\caption{Architecture of the initial SIESTA proposal (white background) along with its extensions in this work (light blue background)}
	\label{fig:architecture}	
\end{figure}

The architecture of the early implementation of SIESTA (illustrated in Figure~\ref{fig:architecture}) consists of two main components: the pre-processing component and the query processor. The pre-processing component is responsible for handling continuously arriving logs and computing the appropriate indices, while the query processor utilizes the stored indices to perform efficient pattern analysis.


The primary inverted index, named IndexTable, is created based on the \textit{et-pairs}. 
For each \textit{et-pair}, we extract from each trace a list of (non-overlapping in time) \textit{event-pairs} and combine them in a single list, by keeping for each \textit{event-pair} the $trace_{id}$ in which it was detected and the timestamps of the two events in the pair.
In addition to the IndexTable, there are several auxiliary tables that enable different processes. A complete list of all the tables used is presented below, including their purpose and structure. It is important to note that the structure of all the tables follows a key-value pair, as they are stored in Cassandra in \cite{siesta}.

\begin{itemize}
	\item \textbf{IndexTable:} This table is the inverted index and serves as the primary table for answering queries. Its primary key is an \textit{et-pair}, i.e. ($a_i,a_j$), where $a_x \in A$. The structure is $(a_i,a_j)$ : [($trace_{id}$, $ev_k.ts$, $ev_\lambda.ts$), $\dots]$, where $trace_{id}$ uniquely identifies each trace and $ev_k.ts < ev_\lambda.ts $ and $ev_k, ev_\lambda \in t$ with $t$ being the trace with id equal to $trace_{id}$. The inverted list of an \textit{et-pair} ($a_i,a_j$) is denoted as IndexTable[($a_i,a_j$)].
	
	\item \textbf{SequenceTable:} This table is used to store and manage all indexed traces in the system; new events belonging to an already active trace are simply appended to the list. The structure of the table is represented as  $(trace_{id}): [ev_1,ev_2,...,ev_n]$, where $ev_i \in t$ with $t$ being the trace with id equal to $trace_{id}$. 

	\item \textbf{SingleTable:} This table is an inverted index having as key a single event type and facilitates the generation of new \textit{event-pairs} as well as the detection of patterns in group of traces. Its structure is $(a_i): [(trace_{id}, ev_k.ts, ev_k.pos),\dots]$, where $ev_k \in t$ with $t$ being the trace with id equal to $trace_{id}$. The inverted list of an event type $a_i$ is denoted as SingleTable[($a_i$)].
	
	\item \textbf{LastChecked:} This table is used to avoid creating pair of events that have already been created and indexed. The structure of this table is $(a_i,a_j, trace_{id}): ts_{last}$, where for a pair $(a_i,a_j)$ and a trace identifier $trace_{id}$ the last timestamp that this pair is completed is stored, i.e. $ts_{last} = \{ev_j.ts |$ $j \geq \lambda$ $\forall$ ($trace_{id}'$, $ev_k.ts$, $ev_\lambda.ts$) $\in$ IndexTable[($a_i,a_j$)], with $trace_{id}' = trace_{id}$ $\}$. The last timestamp of an \textit{et-pair} ($a_i,a_j$) of a trace with id equal to $trace_{id}$ is denoted as LastChecked[($a_i$, $a_j$, $trace_{id}$)]

	\item \textbf{CountTable:} This table is responsible to store useful statistics for each \textit{et-pair}. The structure is  $(a_i$, $a_j)$ : \emph{[total\_completions,sum\_durations]}, where \emph{total\_completions} is equal to the length of IndexTable[$(a_i$, $a_j)$] and \emph{sum\_durations} is equal to $\sum (ev_\lambda.ts - ev_k.ts) \forall$ ($trace_{id}$, $ev_k.ts$, $ev_\lambda.ts$) $\in$ IndexTable[($a_i,a_j$)].
	
\end{itemize}	

\begin{figure}[tb!]
	\centering
	\includegraphics[width=.65\columnwidth]{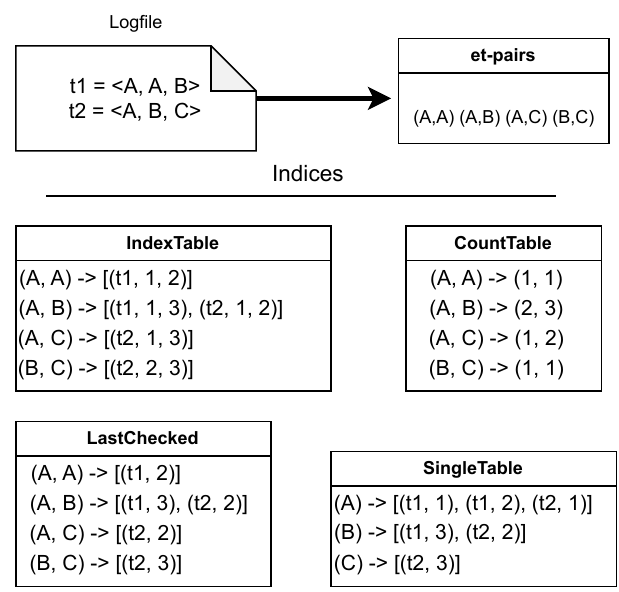}
	\caption{Example of  indices built}
	\label{fig:example_indexing}
\end{figure}

\emph{Example:} Let us consider a web logging application that monitors the order of different actions denoted as A, B, and C, for two users. Figure~\ref{fig:example_indexing} presents this logfile along with the corresponding \textit{et-pairs} and SIESTA's built indices. For simplicity, we use only the position of the events in the traces instead of their timestamps.


Once the indices have been calculated and stored, the query processor utilizes them to respond to different query types. The query input $q$ in the initial SIESTA solution, in all different queries, is a pattern (i.e., a sequence) of $p$ event types $q=< a_1,\dots,a_p>$, where $a_i \in A$. The supported query types in increasing order of complexity are as follows:
\begin{itemize}
	\item \textbf{Statistics.} This query type generates statistics for every consecutive \textit{et-pair} in the pattern, which can provide useful insights about the pattern's behavior through simple post-processing. 
	\item \textbf{Pattern Detection.} This type of query aims to return all traces that contain the given pattern. 
	\item \textbf{Pattern Continuation.} This type of pattern querying returns the events that are most likely to extend the pattern in the query, i.e. it is an exploratory query.
\end{itemize}

To complete the discussion, we provide a brief overview of how the query processor handles pattern detection queries (the remaining query types are not further investigated). For a query pattern of $p$ event types $q=< a_1, \dots,a_p>$, SIESTA uses the IndexTable to detect occurrences of this pattern in the indexed traces. An occurrence of the pattern in a trace $t$ is represented as $Occ = (t, <ev_1, \dots, ev_p>)$ and satisfies $ev_{i}.type = a_i$, with $ev_{i} \in t \land a_i \in q$. 
To that end, query processor first extracts the set of consecutive \textit{et-pairs} from $q$, ($a_i,a_{i+1}$) $\forall i \in [1,p-1]$, and retrieves the corresponding lists stored in the IndexTable. Then, it prunes all traces that do not contain all of the above \textit{et-pairs}, and finally validates these traces to remove false positives, i.e., traces that contain all \textit{et-pairs} but the events do not appear in the correct order. Since the timestamps are already stored in IndexTable, SIESTA can perform the validation without the need to access the original traces in the SequenceTable.

\emph{Example:} Let us consider the small indexed log presented in the Figure~\ref{fig:example_indexing}, and our goal is to detect the pattern $q = <A, B, C>$. First, the set of consecutive \textit{et-pairs} is extracted, which in this case contains the \textit{et-pairs} $(A, B)$ and $(B, C)$. Then, the relevant inverted lists are retrieved from the IndexTable and all the traces that do not appear in both lists are eliminated. As we can observe, only \textit{t2} is present in both lists. Finally, a validation step ensures that \textit{t2} indeed contains the pattern $q$, confirming that it is not a false positive.

\begin{definition}[Occurrences, non-overlapping in time]
	\label{def:non_overlap_ocs}
	Following the Definition~\ref{def:non_overlap}, two occurrences $Occ1=(t_1, <ev_1^1, \dots, ev_p^1>)$ and $Occ2=(t_2, <ev_1^2, \dots, ev_p^2>)$ of a query pattern $q$ of length $p$, with $ev_i^1.type = ev_i^2.type$, do not overlap in time if $ev_p^1.ts < ev_1^2.ts \lor ev_1^1.ts > ev_p^2.ts$.
\end{definition}

It is worth noting that events in each occurrence need not be consecutive, which is enabled by the extraction of \textit{et-pairs} (Definition~\ref{def:event_pair}), making it well-suited for real-world applications where data may contain outliers or unexpected behavior. SIESTA's detection approach is similar to the selection strategy known as Skip-till-Next-Match (STNM) with consume \cite{giatrakos2020vldb} in CEP terminology. In STNM, irrelevant events are skipped until the next matching event is detected. However, in the case of multiple occurrences of the pattern in the same trace, SIESTA extends the traditional definition of STNM by excluding any occurrences that overlap in time (Definition~\ref{def:non_overlap_ocs}). This design choice ensures that the returned information is more compact. 

\section{Query Processor Design}
\label{sec:AQP}

In this section, we introduce the extended query processor (QP) that enhances the capabilities of the original query processor by enabling a wider variety of pattern queries. Due to the improved expressiveness, more sophisticated detections can be performed, such as adding constraints between two events, enabling practical applications like fraud detection and online shopping. Additionally, the processor can now provide explanations about the result set by detecting inconsistencies in both queries and the events' timestamps. These are achieved by integrating SASE into SIESTA. 
In the rest of this section, we provide a generalized formulation of the enhanced pattern detection queries and describe how the various components were implemeneted. These components include time and gap constraints, complex query patterns with Kleene operations and negations, and explanation of non-matches.

\subsection{Pattern Detection Query}
\label{subsec:detection_query}

Following the semantics of the SASE+ language \cite{sase+}, the type of pattern queries supported conform to the following  template:
\begin{tabbing}
	\quad\= FROM $<$process\_name$>$\\
	\> PATTERN $<$pattern q$>$\\
	\> $\small[$WHERE $<$time/gap constraints$>$ $\small]$\\
	\> $\small[$BETWEEN $<$starting time$>$ AND $<$ending time$>$$\small]$\\
	\> $\small[$GROUPS $<$group definition$>$$\small]$\\
	\> $\small[$EXPLAIN-NON-ANSWERS $<k,uncertainty, step>$$\small]$\\
	\> $\small[$RETURN-ALL $<$true/false$>$$\small]$ \\
\end{tabbing}

The query structure above extends the one in the original implementation of SIESTA, which only contained the PATTERN, GROUPS, and RETURN-ALL clauses.  
In some cases, organizations may need to monitor multiple processes separately, such as monitoring the sales and the costumer support. Our comprehensive solution can now support indexing and querying multiple different processes, referred to as log databases.
The FROM clause is used to specify the name of the log database. The same name is also used during indexing to ensure that the new logfiles are correctly appended to the appropriate indices.

The PATTERN clause describes the query pattern. We extend the simpler definition of query pattern in Section ~\ref{sec:background} by allowing CEP operators. That is, the query pattern $q$ is now defined as an ordered list of query events $e$, with $e$ being a tuple $(a, s)$, where $a \in A$ and $s$ is a CEP operator. $s \in [\_, +, *, !, ||]$, thus corresponding to simple event, event with Kleene + operator, event with Kleene * operator, Negation and Or, respectively. 
In the case of an Or operator, we extend the query event with a list of all the alternative event types. For example, if events $A$, $B$, and $C$ are possible alternatives for the position $i$ of a query $q$, then this is being denoted as $e_i = (A, ||, [B, C])$, with $e_i \in q$. The FROM and PATTERN clauses are mandatory in each pattern detection query. 

In the WHERE clause, time and gap constraints between events are defined. Each constraint is a tuple ($gt$,$wa$,$v$,$i$,$j$), where $gt=$ \textit{gap/time}, $wa=$ \textit{within/atleast}, $v$ is the constraint value, and $i$,$j$ denote the positions of the two events in the query pattern. Therefore, for a query pattern $q$ with $|q| = p$, the constraint (\textit{time}, \textit{within}, $value$, $a$, $b$) denotes that for every detected occurrence $Occ = (t, <i_1, \dots, i_p>)$, the event in position $a$ and the event in position $b$ should occur within a timeframe that is less or equal than the given value, $ev_{i_b}.ts - ev_{i_a}.ts \leq value$.
It would have been $\ge$ if, instead of \textit{within}, it was \textit{atleast} and the $pos$ would have been used instead of $ts$ in case of a \textit{gap} constraint. This is especially valuable in applications like fraud detection. For instance, consider a scenario where a pattern $q$ consisting of $p$ events ($p\ge 3$) needs to be detected, but the first three events should occur within a two-minute timeframe. In this case, the WHERE clause will be set to (\textit{time}, \textit{within}, \textit{2}, \textit{1}, \textit{3}).


The BETWEEN-AND clause defines the time window of interest in which we want the matches to have taken place. That is, for a query pattern $q$, with $|q| = p$ and the clause BETWEEN $ts_a$ AND $ts_b$, every detected occurrence $Occ = (t, <i_1,\dots,i_p>)$ should occur within this timeframe, i.e. $ev_{i_1}.ts \ge ts_a \land ev_{i_p}.ts \le ts_b$. Therefore, if it is assumed that a security breach occurs outside of regular working hours, the time window of interest can be adjusted accordingly.

The GROUPS clause allows for custom grouping between traces to be defined during the query phase. Groups are defined as a list of sets, where each set can contain a single number or a range of numbers. For example, the group definition $[(1-3,8),(5-7)]$ represents two groups: one containing the traces with ids $(1,2,3,8)$ and the other containing the traces with ids $(5,6,7)$. This supports scenarios like web logging, where multiple sessions can be grouped based on the IP or an external user id. The implementation of groups has been explained in \cite{siesta} and thus further details are omitted here.

The EXPLAIN-NON-ANSWERS clause is used to enable the detection of both inconsistencies in the query and imprecise timestamps in the traces and it will be explained in detail in Section~\ref{subsec:whyNotMatch}. Note that this option is currently only allowed in simple query patterns $q$ (i.e. $\forall e \in q | e.s = \_ $).



Finally, by setting the RETURN-ALL parameter to true (default is false), multiple non-overlapping occurrences of the query pattern can be returned for each individual trace.  Otherwise, only the first occurrence of the pattern for each trace will be returned to speedup result generation.

\subsection{Handling Gap/Time-Constraints}
\label{subsec:constraints}

As mentioned in Section~\ref{sec:background}, the pattern detection queries are handled in the following way: first, it extracts all the consecutive \textit{et-pairs} that exist in the query pattern, then it retrieves the corresponding inverted lists from the IndexTable, pruning all the traces that do not contain all the pairs, and finally validating the remaining ones based on the retrieved events. However, this approach does not allow for straightforward integration of constraints between events.


For example, consider a trace $t = <A_1, A_2, B_1, C_1>$, where $A_1$ occurs five hours prior to $B_1$, and $A_2$ occurs only thirty minutes before $B_1$. In this case, correctly identifying instances of the pattern $q = < (A,\_), (B,\_), (C,\_) >$ with constraint (\textit{time}, \textit{within}, \textit{60 minutes}, 1, 2) becomes challenging. That is because during indexing, since the event-pairs are not overlapping, there would be one instance of the $(A,B)$ et-pair, namely $(A_1, B_1)$, and one instance for the $(B,C)$ et-pair, namely $(B_1, C_1)$. Therefore, when detecting query $q$, only the events $A_1$, $B_1$, and $C_1$ will be retrieved from the IndexTable, leading to the conclusion that there is no instance of this pattern in $t$. However, the sequence $<A_2, B_1, C_1>$ actually constitutes a valid occurrence of the pattern. Thus, the challenge lies in correctly applying constraints and efficiently retrieving all necessary information from the database to effectively prune the traces before proceeding to the validation step.

\begin{lemma}
	\label{lemma:retrieve_events}
	If a trace $t$ contains multiple events of type $a_i$, then including an \textit{et-pair} ($a_i$, $a_i$) in the list of \textit{et-pairs} to be queried from the IndexTable guarantees the retrieval of all events with that type.
\end{lemma}
\textbf{Proof} The proof is by contradiction. Assume the inverted list IndexTable[($a_i,a_i$)] has been retrieved, but an event $ev_x$ of a trace $t$ with $ev_x.type = a_i$ is not in the list. Also assume that $|o_{a_i}| \ge 2$, where $o_{a_i}$ = [ $ev_j |$ $\forall ev_j \in t$, $ev_j.type = a_i$] and $o$ is sorted based on the timestamps ($ev_x \in o_{a_i}$). Since the \textit{event-pairs} in the IndexTable are generated based on Definition~\ref{def:non_overlap} with its discussed exception when the event types are the same, an \textit{event-pair} ($ev_j,ev_{j+1}$) will be generated $\forall j \in [1, |o_{a_i}|-1]$. Therefore, $ev_x$ will be included in at least one of the indexed \textit{event-pairs} in the IndexTable, which contradicts the assumption that it will not be retrieved. 

Our proposal is based on Lemma~\ref{lemma:retrieve_events}. For a query pattern $q = <(a_1,s_1), (a_2, s_2), \dots >$ with a list of constraints, defined as tuples ($gt$,$wa$,$v$,$i$,$j$), we extend the \textit{et-pairs} to be queried from the IndexTable using the following procedure: (1) if $wa=$ \textit{within} we add ($a_i,a_i$) whereas (2) if $wa=$\textit{atleast} we add an \textit{et-pair} ($a_j,a_j$). We consider these additional \textit{et-pairs} as optional, and so we do not prune traces that do not contain them. The type of the constraint (\textit{gap/time}) is not relevant in this procedure.


\begin{theorem}[Correctness of our proposal]
	\label{theorem:correctness}
	There are no false negatives for any constraint definition, i.e. there are no instances of the query pattern that are mistakenly excluded.
\end{theorem}

\textbf{Proof} If we prove that Theorem~\ref{theorem:correctness} holds for every two consecutive events then it can be generalized for every two pairs in the pattern.  
Let us consider a small query pattern $q = < (a_1, s_1)$, $(a_2, s_2) >$ with a constraint defined as tuple ($gt$,$wa$,$v$,$i$,$j$). ($a_1, a_2$) is the only one \textit{et-pair} in $q$. The IndexTable[($a_1, a_2$)] separates the set of traces into two subsets (those that contain this \textit{et-pair} and those that do not). The latter subset can be safely pruned, as these traces do not contain any \textit{event-pair} ($ev_{i_i}, ev_{i_2}$), where $ev_{i_i}.type = a_i \land ev_{i_2}.type=a_2 \land ev_{i_1}.ts < ev_{i_2}.ts \land ev_{i_2}.ts-ev_{i_1}.ts < lookback$.  

The remaining traces can be further divided into two categories: those that have at least one retrieved \textit{event-pair} satisfying the constraint, and those that do not. The traces in the first category are true positives and are added to the result set. However, we cannot dismiss the traces in the second category yet. For example, if the returned \textit{event-pair} is ($ev_a, ev_b$) and the $wa =$ \textit{within}, there might exist an event $ev_x$ with $ev_x.type = a_1$ such that $ev_a.ts < ev_x.ts < ev_b.ts$ and $ev_b.ts - ev_x.ts \le v$. To retrieve all events of type $a_1$ we have included the \textit{et-pair} ($a_1, a_1$) in the queried pairs (as shown in Lemma~\ref{lemma:retrieve_events}).  Note that if the trace does not have multiple instances of $a_1$, then there can be no other event between the pair in the trace, and so the trace can be safely pruned. Additionally, we do not need to retrieve all events of type $a_2$, since it is guaranteed (by Definition~\ref{def:non_overlap}) that $ev_b$ is the closest event of type $a_2$ after  $ev_a$.

For the case where $wa=$\textit{atleast}, it is possible that an event $ev_y$ with $ev_y.type = a_2$ exists such that $ev_y.ts>ev_b.ts$ and $ev_y.ts - ev_a.ts \ge v$. In this case, we include the \textit{event-pair} ($a_2,a_2$) in the queried pairs. This time, there is no need to retrieve all the events of type $a_1$ as it is guaranteed (by Definition~\ref{def:non_overlap}) that $ev_a$ is the furthest event of type $a_1$ from $ev_b$ (and does not overlap with a previous \textit{event-pair}). If there exists an $ev_a'$ with $ev_a'.type = a_1$ and $ev_a'.ts < ev_a.ts$, then it would be a part of another \textit{event-pair} of ($a_1,a_2$) and therefore would be available for the validation process.
Overall, in every case all the required events are retrieved from the IndexTable and thus there are no false negatives. 

\subsection{Additional operators and SASE integration}
\label{subsec:sase_integration}

In this section, we discuss the implementation of additional operators, such as Kleene+ and negation, in the pattern detection query. As with the previous section, to effectively prune traces and remove false positives, we will ensure that all necessary events are retrieved from the database, by extending the \textit{et-pairs} set. Note that only the original \textit{et-pairs} are used to prune the search space. The additional \textit{et-pairs} are solely responsible for retrieving extra information required for trace validation and, therefore, do not introduce any false positives.
To this end, we define 2 sets of \textit{et-pairs}, $\mathcal{E}_t$ and $\mathcal{E}_a$, with $\mathcal{E}_t \subseteq \mathcal{E}_a$. The former contains the \textit{et-pairs} that should be present in a trace to proceed with the validation step, while the latter contains all the \textit{et-pairs} that will be retrieved from the IndexTable.

\begin{algorithm}[tb!]
\small
	\caption{Calculate $\mathcal{E}_t$ and $\mathcal{E}_a$ }
	\label{alg:etpair_extraction}
	\hspace*{\algorithmicindent} \textbf{Input} Pattern $q= [ e_1, e_2,\dots , e_n]$, $List<Constraints>$ clist\\
	\hspace*{\algorithmicindent} \textbf{Output} $\mathcal{E}_t$, $\mathcal{E}_a$
	\begin{algorithmic}[1]
		\State \textit{qp} $\leftarrow$ \textbf{extract\_different\_queries}(q)
		\State $\mathcal{E}_t$, $\mathcal{E}_a$ $\leftarrow$ \{ \}, \{ \}
		\For{j $\in [1,\dots, m]$, where $m=|qp|$}
		\State $\mathcal{E}_{x_j}$ $\leftarrow$ \{ \}
		\ForAll{$e_i$ in $qp_j$}
		\If{$e_i.s$ $\in$ $["!", "*"]$}
		\State $\mathcal{E}_a$.add( $(e_i.a, e_i.a)$)
		\State $qp_j$.remove($e_i$)
		\EndIf
		\EndFor
		\For{$i \in [1,\dots,|qp_j|-1]$}
			\State $\mathcal{E}_{x_j}$.add($e_i.a,e_{i+1}.a$)
		\EndFor
		\ForAll{$e_i \in qp_j$}
			\If{$e_i.s$ = "+"}
				\State $\mathcal{E}_a$.add($(e_i.a, e_i.a)$)
			\EndIf
		\EndFor		
		\ForAll{($qt$,$wa$,$v$,$i_1$,$i_2$) in clist}
			\If{$wa$ = \textit{within}}
				\State $\mathcal{E}_a$.add($(e_{i_1}.a,e_{i_1}.a)$)
			\Else
				\State $\mathcal{E}_a$.add($(e_{i_2}.a,e_{i_2}.a)$)
				\EndIf
			\EndFor
		\EndFor
		\State $\mathcal{E}_t \leftarrow \mathcal{E}_{x_1} \cap \mathcal{E}_{x_2} \cap \dots \mathcal{E}_{x_m}$
		\State \Return $\mathcal{E}_t$, $\mathcal{E}_a$
	\end{algorithmic}
\end{algorithm}

For a query pattern $q=< e_1, e_2,\dots , e_n>$ and a list of constraints,  Algorithm~\ref{alg:etpair_extraction} generates  $\mathcal{E}_t$ and $\mathcal{E}_a$. First, the algorithm extracts all the possible patterns, i.e. the different patterns that are formed due to the Or ("$||$") operator (line 1). Note that after this operation, there will be no events $e_i$, with $e_i.s = "||"$. For each one of the $j$ possible patterns derived, we perform 3 operations. 
First, for every event with negation or Or operator (i.e. $\forall e_i \in qp_j |e_i.s \in ["!", "*"]$) it adds an $(e_i.a, e_i.a)$ pair in the $\mathcal{E}_a$ and removes it from the $qp_j$, because these events are not required to be part of a trace in order to contain a valid occurrence of the $q$ (lines 6-8).
Next for every two consecutive events in the remaining $qp_j$, the algorithm adds an \textit{et-pair} to the $\mathcal{E}_{x_j}$, as in the original SIESTA QP solution. This set contains the true \textit{et-pairs} for each $qp_j$ (lines 9-10) and the overall set of true \textit{et-pairs} is calculated as the intersection of all $\mathcal{E}_{x_j}$ (line 19). 
Additionally, for each $e_i$ with Kleene $+$ ($e_i.s = "+"$), we add an $(e_i.a,e_i.a)$ pair to the $\mathcal{E}_a$ based on Lemma~\ref{lemma:retrieve_events}(lines 11-13).
Finally, for completeness, lines 14-18 describe the procedure for handling constraints, as represented in the previous section.

After retrieving from the IndexTable all the \textit{et-pairs} in $\mathcal{E}_a$ and prune the traces based on the $\mathcal{E}_t$ contents, the remaining traces proceed to validation. In the original implementation, of the query processor, a linear parser was utilized for that purpose. However, now that the queries contain Kleene closure and negation operators, along with constraints, a more sophisticated approach is required. Modern CEP engines fit that purpose. CEP engines can parse a stream of events and enumerate all the appearances of a predefined pattern based on a selection strategy. Therefore, we integrate SASE~\footnote{\url{https://github.com/haopeng/sase}} with SIESTA, to perform the final validation step. 

The events are first partitioned based on the trace id and then ordered according to their timestamps.
These sequences of events are then processed by SASE, which is initialized with the query pattern. The occurrences of the pattern detected by SASE are subsequently returned to SIESTA. Even though SASE initially lacked the implementation for Kleene* and Or operators, these functionalities were straightforward to implement. Note that, as shown in  \cite{mavroudopoulos2021sequence}, SASE is not suitable for big data. However, the efficient pruning performed by SIESTA (since traces are aggressively eliminated and only the relevant events are present during validation) allows the SASE overhead to be mitigated and outweighed by the benefits of extended functionality.

\subsection{Explaining non-answers}
\label{subsec:whyNotMatch}
End-users of most systems typically lack direct access to the underlying data model, making it difficult for them to verify the reasoning behind the results returned from a query. This lack of transparency often results in a lack of trust in the tool, leading to end-users becoming disengaged \cite{whyNot}. Previous studies on non-answer explanations for database queries have focused on either query explanation \cite{whyNot,conquer_whynot}, which addresses issues in the query, or data explanation \cite{Huang_2008,Herschel_2010}, which addresses issues in the data.
Recently, in \cite{why_not_match}, a framework is proposed that addresses non-answer explanations for streaming data, covering both aspects of query and data explanation. Inspired by this work, we implemented a similar solution on top of SIESTA to detect inconsistencies in the query patterns and stored data.
Before executing a pattern detection query, we first evaluate its consistency through three steps: (1) verifying that all event types in the query exist in the database, (2) ensuring that all extracted \textit{et-pairs} are present in the database, and (3) confirming that each constraint can be satisfied by at least one \textit{event-pair}. To facilitate these checks, we extend the information stored for each \textit{et-pair} in the CountTable to include its minimum and maximum recorded duration, i.e. time difference between the two events. This extension enables us to efficiently perform all three evaluations.


The data explanation takes place during the validation step, when all required events are present. By treating once again the retrieved events as a (sub-trace) stream, we can employ techniques that detect patterns in streams with imprecise timestamps, such as those presented in \cite{zhang_2013_impreccise} and \cite{zhou2014sequence}. 
For our implementation, we chose the Point-based method described in \cite{zhang_2013_impreccise} due to its simplicity and capability to work with SASE, which we have already integrated with SIESTA. This procedure can be enabled during the query phase .

The Point-based method starts by calculating the possible cases (words). That is, each event $ev$ has an uncertainty interval ($ev.ts \pm uncertainty$), and for each timestamp in this interval, a new event is created; the parameter $step$ is used to define the discrete timestamps in this interval, since time is a continuous variable. Once all events have been generated, they are passed as a stream and evaluated by SASE. The skip-till-any-match selection strategy is utilized to ensure that no modification is missed \cite{zhang_2013_impreccise}. If a match is returned that meets all the constraints, we report the modification.

Similarly to \cite{why_not_match}, if there are multiple modifications found, the one with the minimum cost is preferred. The cost is calculated as the sum of the absolute differences between each event's original and modified timestamp and is efficiently counted as the events are processed by SASE. Additionally, the cost must be less than the user-defined parameter $k$ for a modification to be considered valid. Based on this, we ensure that SASE will terminate any partial matches whose cost has exceeded the parameter $k$. The parameters $k$, $uncertainty$, and $step$ are specified in the pattern detection query in the EXPLAIN-NON-ANSWERS clause.

\begin{figure}[tb!]
	\centering
	\begin{subfigure}[t]{0.5\columnwidth}
		\centering
		\includegraphics[width=\columnwidth]{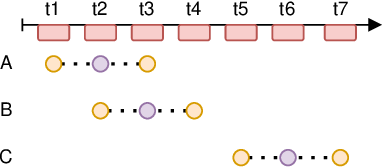}
		\caption{Possible cases (words)}
		\label{fig:whyNotMatchExample}	
	\end{subfigure}%
\begin{subfigure}[t]{0.5\columnwidth}
	\centering
	\includegraphics[width=\columnwidth]{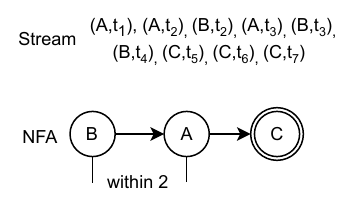}
	\caption{Sase evaluation}
	\label{fig:whyNotMatchExample2}	
\end{subfigure}
\caption{Example with modified timestamps}
\end{figure}

\emph{Example:} Consider an industrial environment with location sensors installed, monitoring the movement of workers. Each time a worker enters a room, the location sensor transmits this information with the worker's id. However, due to possible transmission delays of up to 1 second, the recorded timestamps may not precisely match the actual event times. The administrator wants to detect if a specific sequence $B,A,C$ is followed by any of the workers. The constraints are that the transition from room A to room C should occur within 2 seconds, and considering the possible delays, it allows searching for modified timestamps that sum up to 4 seconds from the recorded timestamps. Based on the above the submitted query will be the following:
\begin{tabbing}
	\quad\= FROM $<$main$>$\\
	\> PATTERN $<(B,\_),(A,\_),(C,\_)>$\\
	\> WHERE (\textit{time}, \textit{within}, 2, 2, 3) \\
	\> EXPLAIN-NON-ANSWERS (4, 1, 1)
\end{tabbing}
For a worker, the events A, B, and C are retrieved with relevant timestamps $t_2$, $t_3$, and $t_6$, respectively. Based on the uncertainty parameter (1 second), the possible timestamps for $ev_1$ are $t_1$, $t_2$, and $t_3$, and the same applies to the other two events (Figure ~\ref{fig:whyNotMatchExample}).  This results in a total of $3^3 = 27$ possible cases. However, the generated stream passed to SASE for evaluation contains only a single extended case with $3\times 3 = 9$ events. Figure~\ref{fig:whyNotMatchExample2} illustrates the stream and the NFA used for evaluation. Although the sequence $<B,A,C>$ does not appear in the retrieved subtrace, an occurrence of the pattern can be found in the timestamps $<t_2, t_3, t_5>$, and the modification cost is equal to 3, which is less than the threshold $k = 4$; $B$'s timestamp is decreased by 1 and $A$'s is increased by 1 to meet the ordering constraints, while $C$ is also modified (decreased by 1) to meet the \emph{within} constraint. Overall, the non-answer explanation lies in producing this modification to notify the user under which altered timestamps the trace id would have been returned.

\section{Storage}
\label{sec:storage}
In this section, we outline the storage design choices that differ from the initial implementation in \cite{siesta}. 

\subsection{Object Storage}
\label{subsec:object_storage} 
Here, we explore an alternative approach that aims to further increase SIESTA's storage scalability and cost-efficiency, making it suitable for a wider range of applications and easier to deploy by real-world organizations. This alternative utilizes an  Object Storage System (OSS) and in our implementation, we opted for the S3 protocol through MinIO\footnote{\url{https://min.io/}} due to its widespread adoption in both the industry and the research community~\cite{logstore}.
S3 simplifies deployment, costs less and compresses data before transmission. It utilizes parquet files\footnote{\url{https://parquet.apache.org/}} for data storage, which are immutable and can be partitioned based on column values. This immutability poses two main challenges that need to be addressed. 
Firstly, in incremental indexing (discussed in the next section), the process involves fetching the previous records, combining them with the new data, and rewriting them back to S3, making the process more time-consuming than it would be otherwise. Secondly, we must ensure that using S3 in SIESTA does not compromise the system's fast response times, considering that the data is stored in a less optimized index lookup structure compared to Cassandra.

To address these challenges, we have implemented partitions as a solution. We introduced two new parameters during index building: \emph{split\_every\_days} and \emph{trace\_split}. The \emph{split\_every\_days} parameter allows users to specify a time interval in days for splitting the IndexTable and SingleTable into non-overlapping segments. For example, if set to 30, both tables will be split into 30-day segments. Note that in the case of the IndexTable, if a pair of events spans multiple segments, it is added to the segment of the last event. Similarly, the \emph{trace\_split} parameter partitions the SequenceTable and LastChecked based on trace groups. For instance, if set to 10K, a new partition will be created every 10K traces. In addition, the \emph{split\_every\_days} parameter helps to limit the size of the inverted lists and refer to bounded time periods; otherwise they would grow infinitely (for this reason, this feature is implemented in Cassandra as well).

Furthermore, to optimize response times, we have implemented data partitioning based on the original columns. However, partitioning the data using the primary keys that are used in Cassandra tables will sometimes result in an excessive number of partitions, like in the case of IndexTable, negatively impacting times. Therefore, instead of using the entire \emph{et-pair} as the partition key in IndexTable, LastChecked, and CountTable, we have chosen to partition these tables based solely on the first event of the \emph{et-pair}. This approach allows us to retain the benefits of partitioning in response times without compromising the efficiency of the preprocessing procedure. 
For example, the structure of the IndexTable that was previously stored in Cassandra (as shown in Section~\ref{sec:background}) is  transformed to the following format: IndexTable: ($a_i$, $a_j$, $ts_{start}$, $ts_{end}$, [($trace_{id}$, $ev_k.ts_a$, $ev_\lambda.ts$), $\dots$]), partitioned first by the time interval ($ts_{start}$, $ts_{end}$) which is derived from the \emph{split\_every\_days} parameter, and then by the first event in the et-pair (i.e., $a_i$).

\subsection{Incremental Indexing}
\label{subsec:incremental_indexing}
Implementing efficient incremental indexing can be challenging, especially for systems like SIESTA that have multiple indices with relatively high building times. Minimizing index building time is critical for maintaining the practicality and real-world applicability of SIESTA. We target scenarios where indices are updated on a daily or an hourly basis or even more frequently, e.g., every 15 minutes.

The biggest challenge is to efficiently handle the traces that expand in multiple log batches. The additional events belonging to an ongoing trace require combining them with the previous events in order to generate all the corresponding \textit{event-pairs}. Consider a scenario where we log user events on a website and create a new logfile every day at midnight. In this scenario, for example, a user logs in at 23:50 and performs the events $A_1$, $B_1$, and $A_2$ until midnight. After midnight, the user is still browsing the website and performs the events $B_2$ and $D_1$, which are logged to a newly created logfile. When the new logfile is processed for indexing at the end of the second day, we expect to see, among others, the \textit{event-pairs} $(A_1,D_1)$ and $(A_2,B_2)$.

Moreover, in certain domains, it is necessary to consider only those \textit{event-pairs} that occur within a specific time frame to improve efficiency. To meet this requirement, we have introduced a final parameter called $lookback$, which is set during the preprocessing stage. This parameter defines the maximum number of days that can separate two events for them to be considered related and create an \textit{event-pair}.
For example, consider a business process logging application where each process can potentially run for months. If the organization sets the $lookback$ parameter to 30, it defines that two events are related if they occur within a month. Therefore, \textit{event-pairs} with a time difference of more than a month will not be stored in the IndexTable, reducing the size of the main index and improving the query performance.

The key ideas for an efficient incremental processing are to (1) avoid creating the pairs that have already been created and indexed, and (2) effectively update the various tables, while also ensuring that all constraints, such as the time frame between events, are met.

\begin{algorithm}[tb!]
\small
	\caption{Extract Pairs}
	\label{alg:extract_pairs}
	\begin{algorithmic}[5]
		\Procedure{Extract}{$lookback$, $split\_every\_days$,traces}
			\State newPairs $\leftarrow$ empty list
			\For{every $trace$ \textbf{in} traces}
				\State comb $\leftarrow$ create all \textit{et-pairs}
				\For{every ($a_i,a_j$) \textbf{in} comb}
					\State $ts_{list}a_i$ $\leftarrow$ $trace$.getAllTsOf($a_i$)
					\State $ts_{list}a_j$ $\leftarrow$ $trace$.getAllTsOf($a_j$)
					\State $ts_{last}$ $\leftarrow$ LastChecked[($a_i$,$a_j$,$trace.id$)]
					\State pairs $\leftarrow$ \textbf{\textsc{Extract\_Pairs}}($ts_{list}a_i$, $ts_{list}a_j$, $ts_{last}$, $lookback$)
					\State newPairs.append(pairs)
				\EndFor
			\EndFor
			\State inter $\leftarrow$ createIntervals(newPairs, $split\_every\_days$)
			\For{every p \textbf{in} newPairs}
				\State p.assignInterval(inter)
			\EndFor
			\Return newPairs			
		\EndProcedure
		\Procedure{extract\_pairs}{$ts_{list}a_i$, $ts_{list}a_j$, $ts_{last}$, $lookback$}
		\If{$ts_{last}$ is not null} 
			\State remove $ts | ts < ts_{last}$, $ts \in ts_{list}a_i \lor ts \in ts_{list}a_j$
		\EndIf
		\State pairs $\leftarrow$ empty list
		\For{every $ts_i$ \textbf{in} $ts_{list}a_i$}
				\State find the first $ts_j$ in $ts_{list}a_j$ so
				\State (a) $ts_i \leq ts_j$
				\State (b) $ts_j$ hasn't been used as second event
				\State (c) $ts_j$-$ts_i$ $\leq$ $lookback$
				\State pairs.append($ts_i$,$ts_j$)
		\EndFor
		\Return pairs
		\EndProcedure
	\end{algorithmic}
\end{algorithm}

For a newly arrived logfile, the preprocess component will start by calculating the single inverted index for the new events. Then, it will update SequenceTable and SingleTable with the new information. Next, the updated traces from the SequenceTable, along with the parameters $lookback$ and \emph{split\_every\_days}, will be used by the incremental indexing procedure, outlined in Algorithm\ref{alg:extract_pairs}, to extract the new \textit{event-pairs}.

For each trace all the \textit{et-pairs} are generated. Then, for each \textit{et-pair} we extract a list of the timestamps of each one of the two event types in the pair (lines 6-7). That is, the function \textit{getAllTsOf}, when called for a trace $t$ and an event type $a_i$, it will return a list of all the events with that type in  $t$ (i.e. $[ev.ts|$,  $ev \in t, ev.type = a_i]$). Next, to avoid recalculating already indexed \textit{event-pairs}, the algorithm retrieves the last timestamps of each \textit{event-pair} per trace from the LastChecked (line 8). The two lists with the timestamps that correspond to event types $a_i$ and $a_j$ along with the last timestamp and the $lookback$ parameter are passed to the \textsc{Extract\_Pairs} procedure to generate the \textit{event-pairs}.

\textsc{Extract\_Pairs} begins by pruning all the timestamps in both lists that are before the $ts_{last}$. Then, it parses through the timestamps in $ts_{list}a_i$ and for each $ts_i$, it searches for a $ts_j$ from the $ts_{list}a_j$ that meets specific criteria: (a) it occurs after the former one, (b) it has not been used in a previous pair as second timestamp, and (c) it occurs within a timeframe former the first one, defined by the parameter $lookback$.  When $a_i=a_j$, it holds that we can reuse $a_j$.
Once all the pairs for each different key have been generated, the procedure creates a list of time intervals (non-overlapping time-windows with a length defined by the parameter \emph{split\_every\_days}), and assigns each pair to one interval, based on the timestamp of the second event in the pair (lines 12-14).

\subsection{Storage Optimizations}
\label{subsec:storage_optimizations}
Minimizing the storage space also has a significant impact on response times, as the amount of data that needs to be transferred from the database to the QP in order to answer a query pattern is significantly reduced.
The first step to optimize the storage space is to store the positions of the events in a pair within a trace, instead of their exact timestamps. This simple change resulted in a significant reduction in storage space, as we replaced two strings of approximately 20 characters with two integers. 
The overall index building process remains unchanged, as both timestamps and positions are available during preprocessing. For querying, there are only a few modifications since the positions of events are mainly used to evaluate the occurrence of a pattern. However, in cases where a pattern detection query involves time constraints, necessitating time information, the original trace must be retrieved from the SequenceTable. 
Finally, to further optimize storage space, we have implemented support for several compression algorithms in SIESTA. These algorithms include Snappy~\cite{snappy}, LZ4~\cite{lz4}, and ZSTD~\cite{zstd}, which are supported by both S3 and Cassandra. By using these algorithms, we can compress the data stored in the indices, thereby reducing the amount of storage required. 

\section{Evaluation}
\label{sec:evaluation}


{\bf Datasets.}
We selected the datasets from the Business Process Intelligence (BPI) Challenges of 2017, 2018, and 2019~\cite{bpi_2017,bpi_2018,bpi_2019}, as they contained the largest number of events compared to other available datasets. In addition, we created a synthetic dataset with 10,000 traces and 1 million events spanning 150 different event types. The key characteristics of these 4 datasets are presented in Table~\ref{table:datasets}. Note that the synthetic dataset differs from the real-world ones in that activities were randomly selected from a uniform distribution, unlike the real-world datasets, which follow a power-law distribution.


\begin{table}[tb!]
\centering
\begin{tabular}{|l|r|r|r|r|}
\hline
 & \multicolumn{1}{l|}{\textbf{BPI\_2017}} & \multicolumn{1}{l|}{\textbf{BPI\_2018}} & \multicolumn{1}{l|}{\textbf{BPI\_2019}} & \multicolumn{1}{l|}{\textbf{Synthetic}} \\ \hline
\textbf{Events}                 & 1,202,267                               & 2,514,266                               & 1,595,923                               & 1,000,000                               \\ \hline
\textbf{Traces}                 & 31,509                                  & 43,809                                  & 251,734                                 & 10,000                                  \\ \hline
\textbf{Event Types}            & 26                                      & 41                                      & 42                                      & 150                                     \\ \hline
\textbf{Mean}                   & 38.1                                    & 57.3                                    & 6.3                                     & 100                                     \\ \hline
\textbf{Min}                    & 10                                      & 24                                      & 1                                       & 50                                      \\ \hline
\textbf{Max}                    & 180                                     & 2973                                    & 990                                     & 150                                     \\ \hline
\textbf{Distribution}           & power-law                               & power-law                               & power-law                               & uniform                                 \\ \hline
\end{tabular}
\caption{Dataset characteristics}
\label{table:datasets}
\end{table}

{\bf Competitors.}
We evaluate our comprehensive implementation of SIESTA, containing the extended query processor, utilizing both Cassandra and S3, against ELK  and two other index-based systems, namely Signatures \cite{nanopoulos_2002} and Set-Containment \cite{lcjoin}, that can support arbitrary pattern detection over big data. 
In our evaluation, we use ELK version v8.7.0, which supports a wider variety of pattern detection queries than the previous versions.
Signatures and Set-Containment are the implementations of ~\cite{nanopoulos_2002} and ~\cite{lcjoin}, respectively, on top of the SIESTA framework. The code for these implementations, both for the index building and the integration with the query processor, is publicly available\footnote{\url{https://github.com/mavroudo/SequenceDetectionPreprocess/tree/v2.1.1}, \\\url{https://github.com/mavroudo/SequenceDetectionQueryExecutor/tree/v2.0.0}}.

Furthermore, we assess the performance of two more systems, namely FlinkCEP and MATCH \_RECOGNIZE (MR), during the query phase. We employed FlinkCEP version 1.17.1 and the Trino implementation of MR (version 429) \cite{match_rec_trino}.

\subsection{Incremental Indexing}
\label{subsec:incrementa_indexing}
In the first set of experiments, we evaluate the performance of incremental indexing. We test all the datasets mentioned above and modify each logfile to last for one day, i.e. the difference between the first event in the logfile and the last event being less than 24 hours. Next, we change the trace ids so that 10\% of the traces from one logfile continue to the logfile of the next day, simulating a real-world scenario where millions of events are indexed every day.
All experiments were conducted on a small cluster comprising five machines: one with 64GB of RAM and 6 cores (12 threads) running at 2.1 GHz, and four identical machines with 32GB of RAM and 6 cores (6 threads) running at 2.9 GHz. The network speed between the machines was 100Mbps. For all systems that rely on the SIESTA framework, which will be referring as SIESTA-based from now on, the database (Cassandra/MinIO-S3) and Spark Master both run on the first machine, with 30GB of RAM and 2 cores allocated to the former and 30GB of RAM and 4 cores to the latter. Four Spark workers were deployed on the other four machines, each with full access to all local resources. The ELK system ran in cluster mode on all five machines with complete control of the available resources.

\begin{figure}[tb!]
	\centering
	\begin{subfigure}[b]{0.4\columnwidth}
		\includegraphics[width=\textwidth]{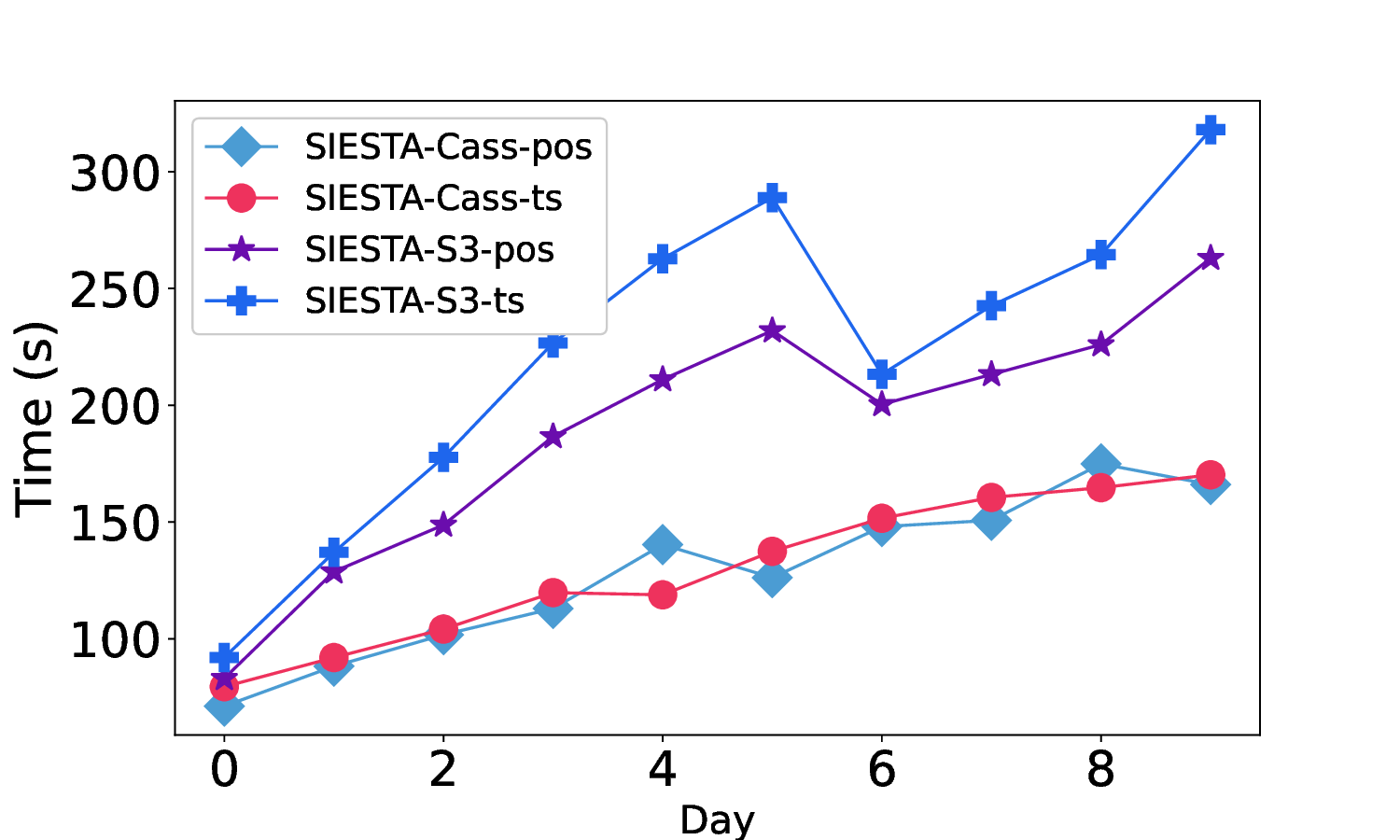}
		\caption{BPI\_2017}
	\end{subfigure}
	\begin{subfigure}[b]{0.4\columnwidth}
		\includegraphics[width=\textwidth]{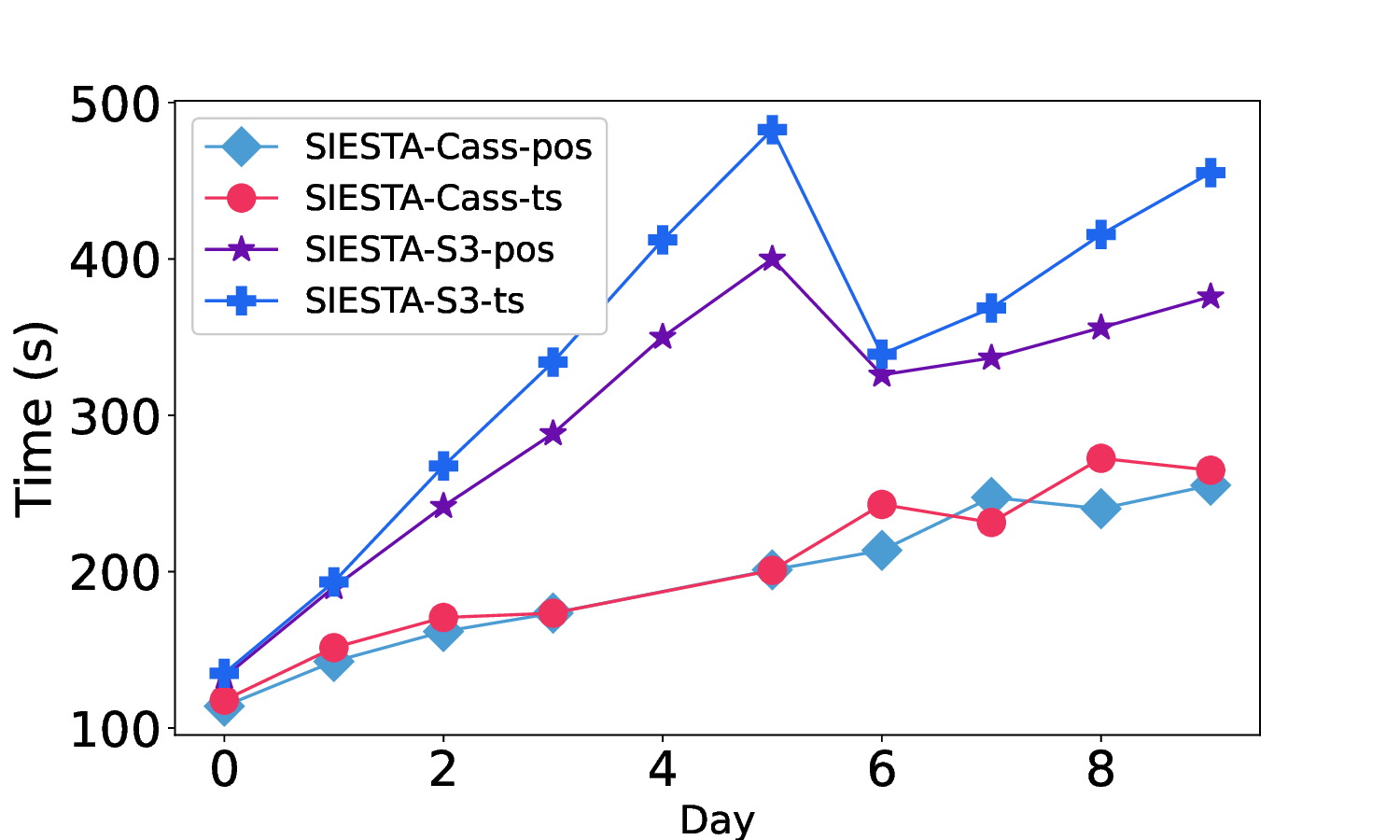}
		\caption{BPI\_2018}
	\end{subfigure}
	\begin{subfigure}[b]{0.4\columnwidth}
		\includegraphics[width=\textwidth]{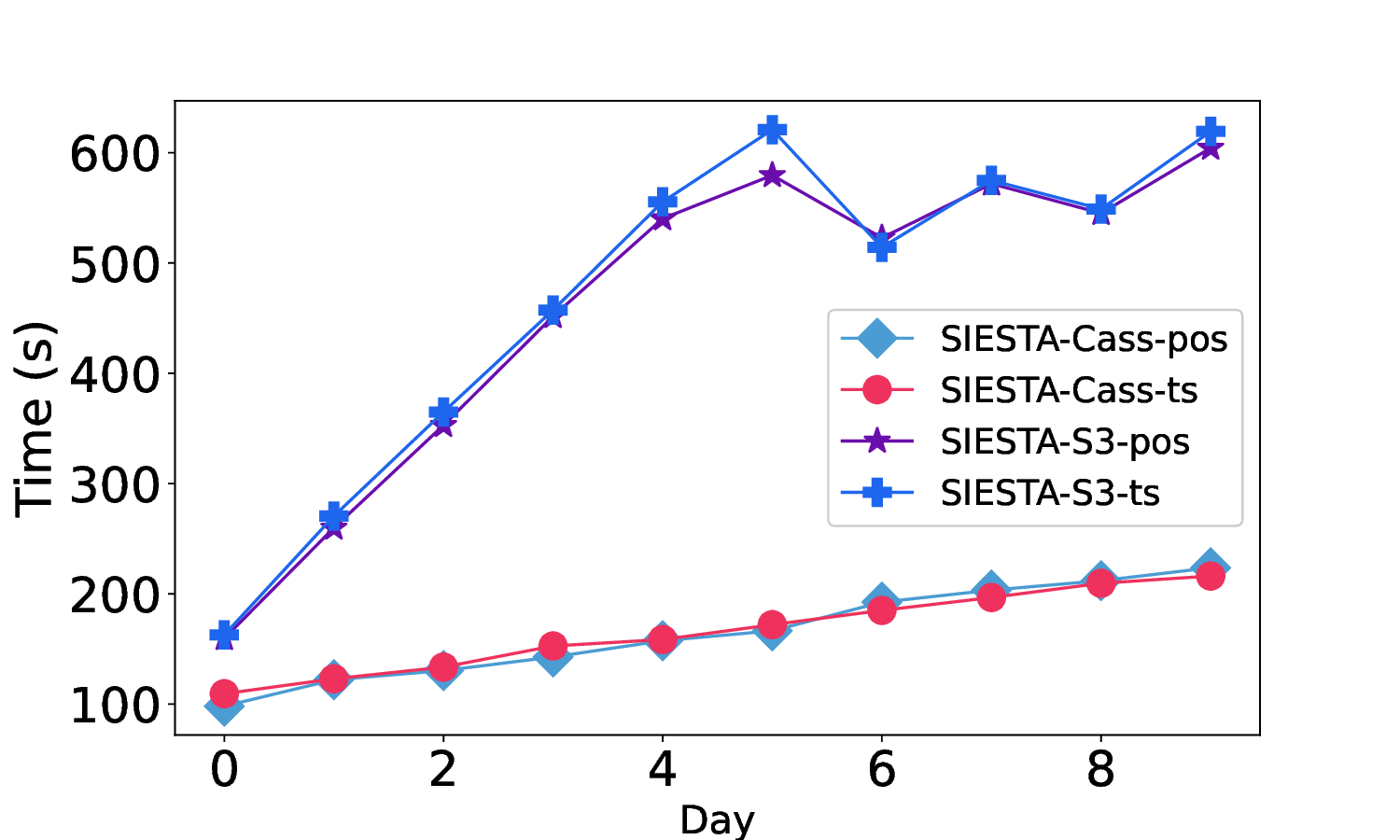}
		\caption{BPI\_2019}
	\end{subfigure}
	\begin{subfigure}[b]{0.4\columnwidth}
		\includegraphics[width=\textwidth]{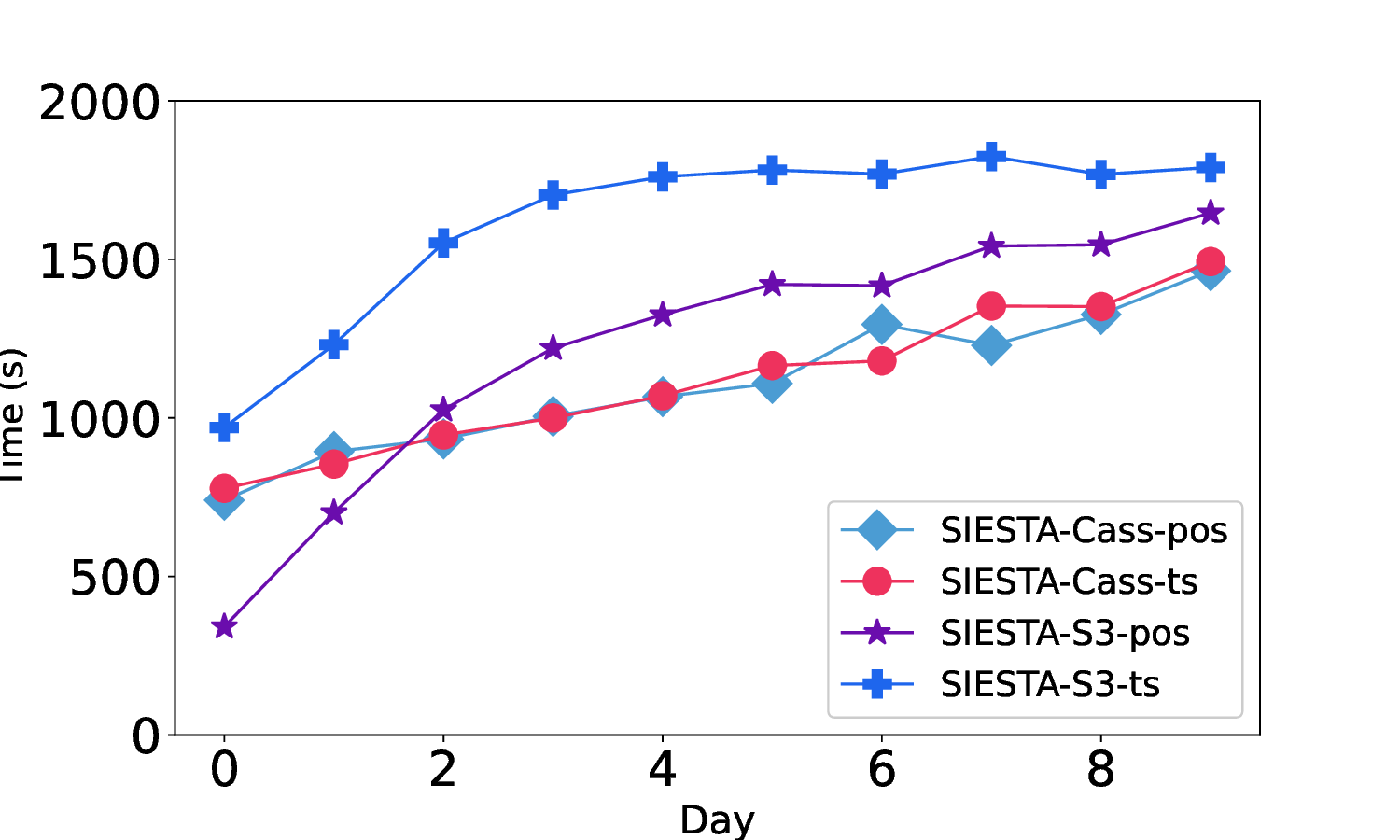}
		\caption{synthetic}
	\end{subfigure}
	\caption{Indexing times for the different datasets during a 10 days period.}
	\label{fig:indexing_ts_vs_pos}
\end{figure}

First, we investigate the impact of using S3 instead of Cassandra for the two different modes of storing the positions (pos) or the timestamps (ts) in the IndexTable. Note for all the experiments, we utilize the Snappy~\cite{snappy} compression algorithm for both databases. The results are presented in Figure~\ref{fig:indexing_ts_vs_pos}.
As demonstrated in all four plots, Cassandra outperforms S3, and storing positions is generally faster than storing timestamps.

More specifically, Figure~\ref{fig:indexing_ts_vs_pos} offers insights into the impact of various parameters on the indexing process. We maintained a consistent 3-day \emph{lookback} period for all experiments. The \emph{split\_every\_days} parameter was set to 5 for the real-world datasets and 30 for the synthetic dataset. That is, the IndexTable was kept as a single table throughout the experiments for the synthetic dataset, while it was split into two tables for the real-world dataset. For S3, we set the \emph{trace\_split} to match the number of traces in each log, ensuring each LastChecked partition maintained records of approximately one dataset.

The key findings are as follows: (1) Cassandra's indexing times consistently increased for all datasets, with minimal influence from the \emph{split\_every\_days} parameter. In contrast, S3 showed promising signs of efficient incremental processing, particularly with shorter inverted lists. In the real-world datasets, S3's indexing time exhibited a drop on the 6th day and barely exceeded the time of the previous (5th) day. This drop occurs because the new segment in the IndexTable is generated on the 5th day, but some \textit{event-pairs} need to be stored in the previous segment, resulting in the observed improvement on the 6th day.
(2) Even though the IndexTable in the synthetic dataset remained unsplit, implying that S3 had to read and recreate the entire table daily, S3 demonstrated nearly constant indexing times after the 4th day, especially when timestamps were used. In contrast, Cassandra's indexing time continued to rise. It is worth noting that the synthetic dataset posed the most demanding scenario for SIESTA, with a larger volume of \textit{event-pairs} generated daily (35 million compared to 4 million in the real-world datasets). These findings suggest that if this trend continues, S3 may eventually outperform Cassandra in terms of indexing.

S3's performance is largely attributed to its efficient compression. The use of immutable parquet files for storing indices enables the preprocessing component to benefit from high compression rates, saving time by minimizing network traffic. However, it is essential to consider other factors such as required storage space, cost-efficiency, and query response times before drawing conclusions.


\begin{table}[tb!]
\small
\centering
	\begin{tabular}{|l|r|r|r|r|}
		\hline
		\textbf{Dataset} &
		\multicolumn{1}{l|}{\textbf{S3-pos}} &
		\multicolumn{1}{l|}{\textbf{Cass-pos}} &
		\multicolumn{1}{l|}{\textbf{S3-ts}} &
		\multicolumn{1}{l|}{\textbf{Cass-ts}} \\ \hline
		BPI\_2017 & \textbf{903} & 1,648 & 1,501 &  2,262\\ \hline
		BPI\_2018 & \textbf{1,320} & 2,580 & 2,002 &  3,331\\ \hline
		BPI\_2019 & \textbf{572} & 1,565 & 638 &  1,904\\ \hline
		synthetic & \textbf{4,920} & 12,249 & 9,709 &  16,129\\ \hline
	\end{tabular}
	\caption{Size in mb at the end of indexing (after 10 days).}
	\label{tab:size_real}
\end{table}

Table~\ref{tab:size_real} displays the required storage space for storing all the indices for each dataset and configuration at the end of the 10th day. The results indicate that S3 significantly benefits from compression, requiring considerably less storage space than Cassandra for the same mode, with space savings of up to 64\%.

To conduct a more detailed cost analysis, we examined Amazon, one of the leading service providers. We aimed to determine the cost an organization would incur to index the synthetic dataset for 10 days using the position of the events and subsequently store the results for a month. We explored the pricing options for both S3 and Keyspaces (equivalent to Cassandra) for a region close to our university. For S3, the organization would pay 0.023\$/GB stored per month. Since data transfers to S3 are free and data transfers from S3 are billed after the first 100GB (after that is 0.009\$/GB), the total cost for the entire process would be 0.092\$ ($4\times0.023$).
For Keyspaces, the billing is based on write request units (WRUs) and read request units (RRUs). A WRU is equivalent to storing one row of data of up to 1KB, while an RRU is equivalent to reading data up to 8KB (LOCAL\_ONE consistency). The costs are 1.48\$ for every million WRUs and 0.29\$ for every million RRUs. The entire process requires 55.7 million WRUs and 4.9 million RRUs, resulting in a total cost of 82.56\$, plus 3.6\$ to store the data for a month (0.3/GB\$ per month). Among the 55.7 million WRUs, LastChecked accounts for 49.5 million WRUs, since it contains a large number of small rows (each row corresponds to a combination of \textit{et-pair} and $trace_{id}$). To reduce the cost, one could argue that instead of using a separate row for each combination, multiple rows could be grouped based on the et-pair, resulting in only 12.3 million WRUs and a total cost of $21.65$ ($17.76$ for WRUs, $0.29$ for RRUs, and $3.6$ for storage). However, this approach would require retrieving and combining data from the LastChecked every time, adding an extensive overhead and resulting in much higher indexing times (detailed results are omitted). 
These findings support the argument that S3 is a more cost-efficient solution for SIESTA, since the savings are on the orders of magnitude. However, it remains to be seen how this affects query response times, which will be evaluated in the next section.

\begin{figure}[tb!]
	\centering
	\begin{subfigure}[b]{0.48\columnwidth}
		\includegraphics[width=\textwidth]{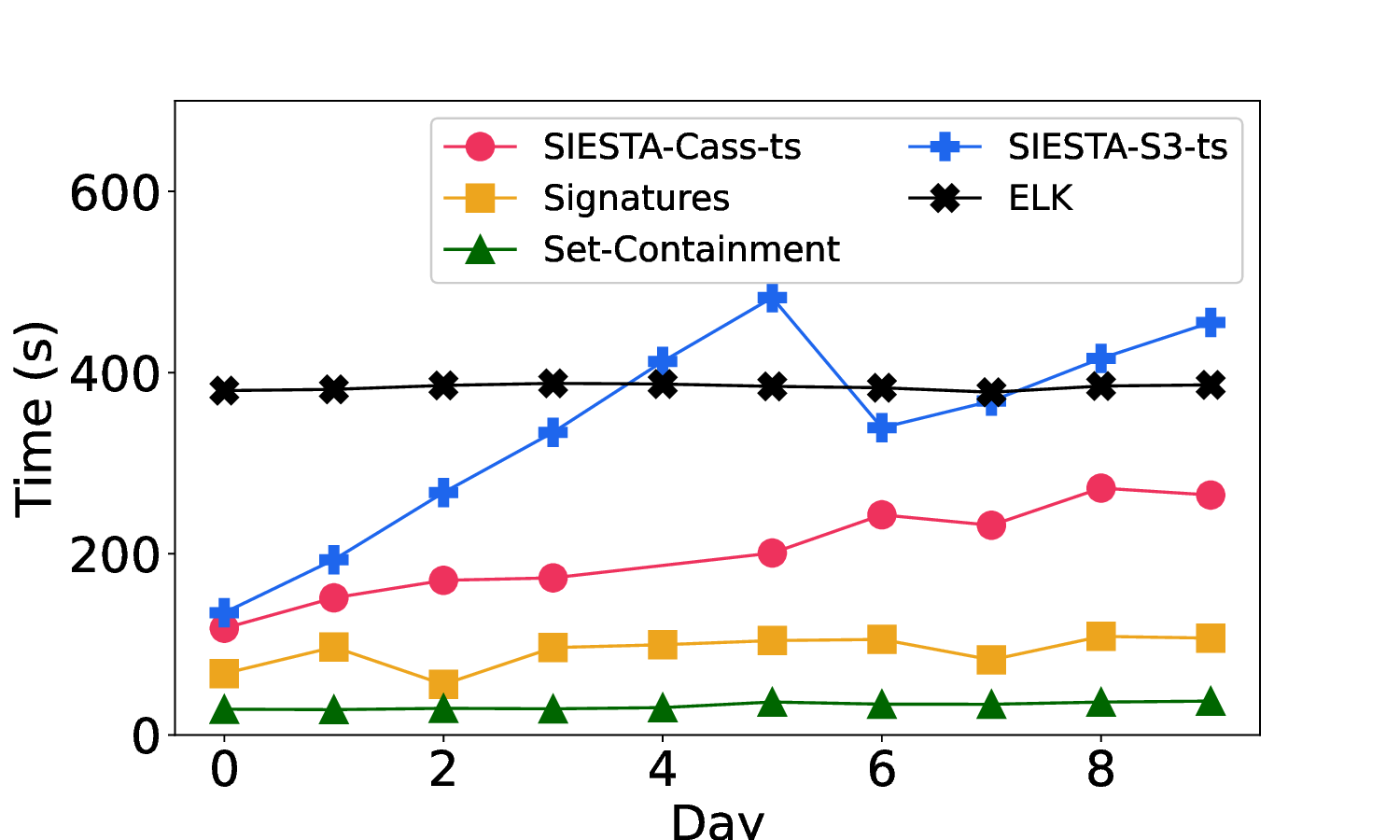}
		\caption{BPI\_2018}
	\end{subfigure}
	\begin{subfigure}[b]{0.45\columnwidth}
		\includegraphics[width=\textwidth]{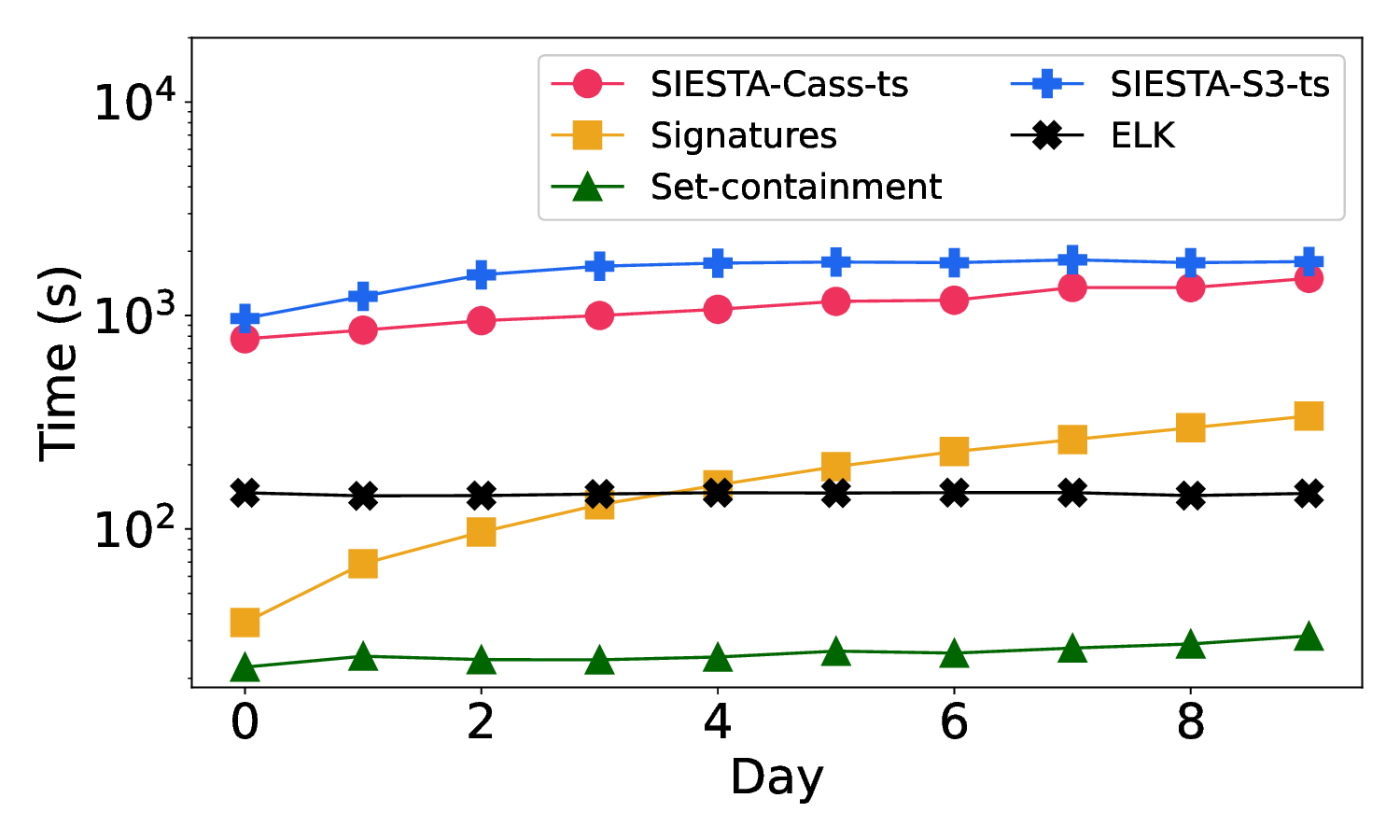}
		\caption{synthetic}
	\end{subfigure}
	\caption{Indexing times for different datasets and various systems.}
	\label{fig:indexing_synthetic}
\end{figure}

In Figure~\ref{fig:indexing_synthetic}, we present the indexing time for the various alternatives when indexing the BPI\_2018 and synthetic datasets. Note that the BPI\_2018 dataset is almost twice as large as the synthetic dataset. In both datasets, Set-Containment achieved the best performance. In the synthetic dataset, where event types follow uniform distribution, SIESTA showed the worst performance, followed by Signatures.
This is because in the uniform distribution, there are more \textit{event-pairs} generated, as most of the events appear only once in each trace, and thus, the number of the generated \textit{event-pairs} is closer to $n(n-1)/2$. On the other hand, in the real-world datasets, where the event types follow a power-law distribution, there may be multiple appearances of the same event type in the trace, which decreases the number of event pairs thus speeding up the indexing time of SIESTA. In the BPI\_2018 dataset, SIESTA outperformed the state-of-the-art system ELK showing faster indexing times, especially when Cassandra is utilized.

As a final remark, we can see that both implementations of SIESTA are applicable in real-world scenarios, as they can effectively build the indices for the real-world data in under 7 minutes for 2.5 million new events for the BPI\_2018 dataset on the 10th day.


\subsection{Query Processor Experiments} 

In this section, we evaluate the response times of the different systems in pattern detection queries. We execute 100 random queries for each experiment, with patterns that appear in at least one trace. The queries were performed on the complete datasets presented in the previous section.
The experiments were conducted on the strongest machine among the five presented before. We allocated 30GB of memory to the query processor of the SIESTA-based systems, while the rest of the memory was reserved for the database. ELK, FlinkCEP, and MR were executed on the same machine and had complete control over the available resources.

\begin{figure}[tb!]
	\centering
		\begin{subfigure}[b]{0.45\columnwidth}
		\includegraphics[width=\textwidth]{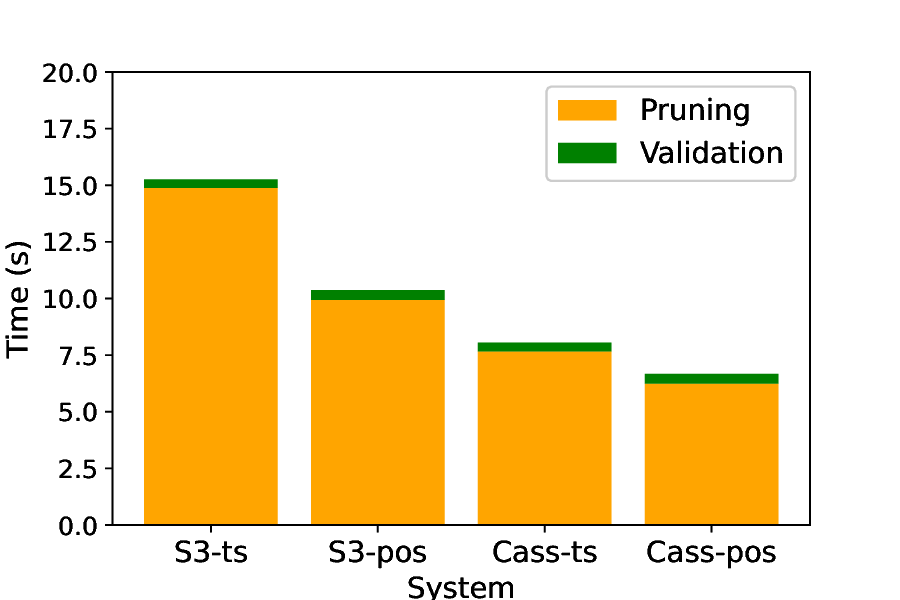}
		\caption{$|q|=2$}
	\end{subfigure}
	\begin{subfigure}[b]{0.45\columnwidth}
		\includegraphics[width=\textwidth]{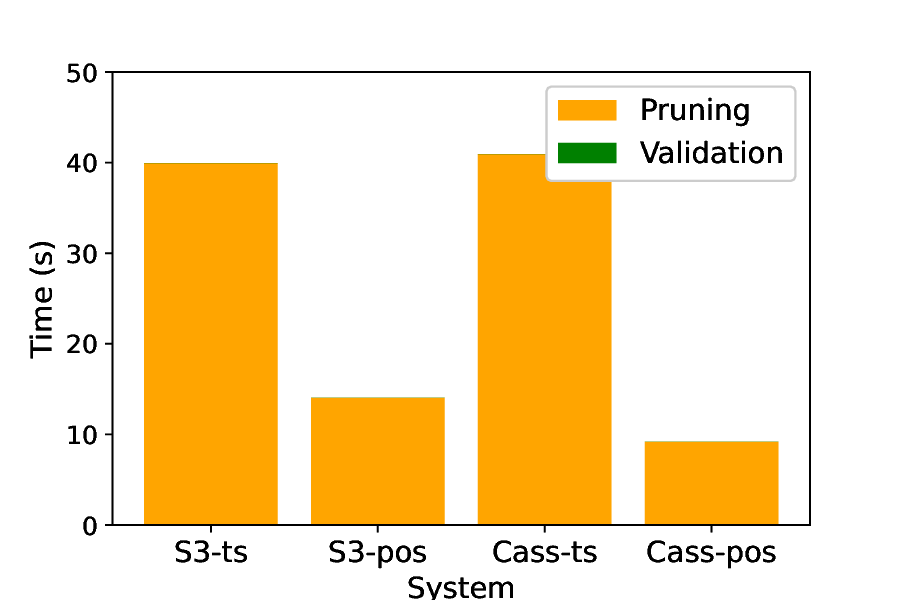}
		\caption{$|q|=10$}
	\end{subfigure}
	\caption{Response times for the SIESTA variations for query lengths equal to 2 and 10.}
	\label{fig:q_ts_pos}
\end{figure}

First, we evaluate the impact of database choice and mode on response times for pattern detection queries using the synthetic dataset. The results, presented in Figure~\ref{fig:q_ts_pos}, show that for smaller queries, Cassandra outperforms S3, while for larger queries, there is no noticeable difference. Storing positions performs better than storing timestamps in all setups, as the overhead of transferring and deserializing \textit{event-pairs} with timestamps outweighs the benefits of not having to retrieve complete traces from the SequenceTable. For larger queries, with more \textit{et-pairs} to retrieve, responses are three times faster. Therefore, we will use positions for the remaining experiments in this section.
Furthermore, we also observed that the majority of time spent in query answering is on fetching and pruning the traces rather than on validation. This suggests that SIESTA is successful in pruning the search space of traces, particularly for queries of length 10, where the retrieved traces are highly accurate and the validation step is completed quickly, taking only a few milliseconds. This effective pruning capability of SIESTA allows it to efficiently incorporate  SASE to support more expressive and complex queries.

\begin{figure}
    \centering
    \includegraphics[width=0.7\columnwidth]{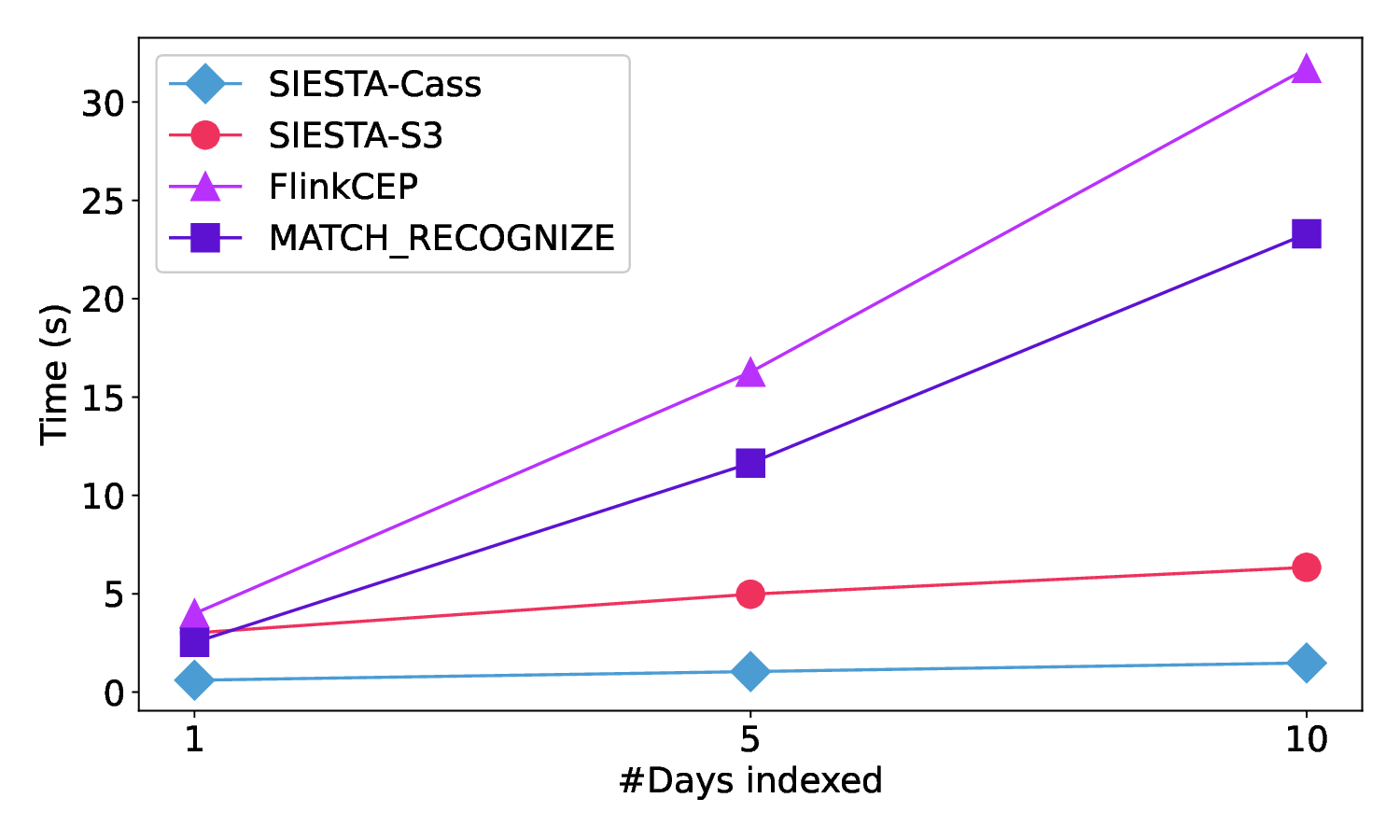}
    \caption{Response times for simple pattern detection queries utilizing the synthetic dataset.}
    \label{fig:compare_static}
\end{figure}

We then proceeded to assess the performance of FlinkCEP and MR in pattern detection across a continuously expanding sequence database. Figure~\ref{fig:compare_static} presents the average response times for 100 queries, of length equal to 5, performed by each system. These queries were conducted using synthetic dataset in three distinct scenarios: (1) employing only the events from the first day, (2) using events from the first 5 days, and (3) utilizing events from the full 10 days. 
Evidently, SIESTA is highly scalable with respect to the data volume, regardless of the utilized database. Furthermore, it is demonstrated that both FlinkCEP and MR struggle with large data volumes due to their lack of an efficient index, necessitating a complete data scan for every query.

\begin{figure*}[tb!]
	\centering
	\begin{subfigure}[b]{0.45\columnwidth}
		\includegraphics[width=\textwidth]{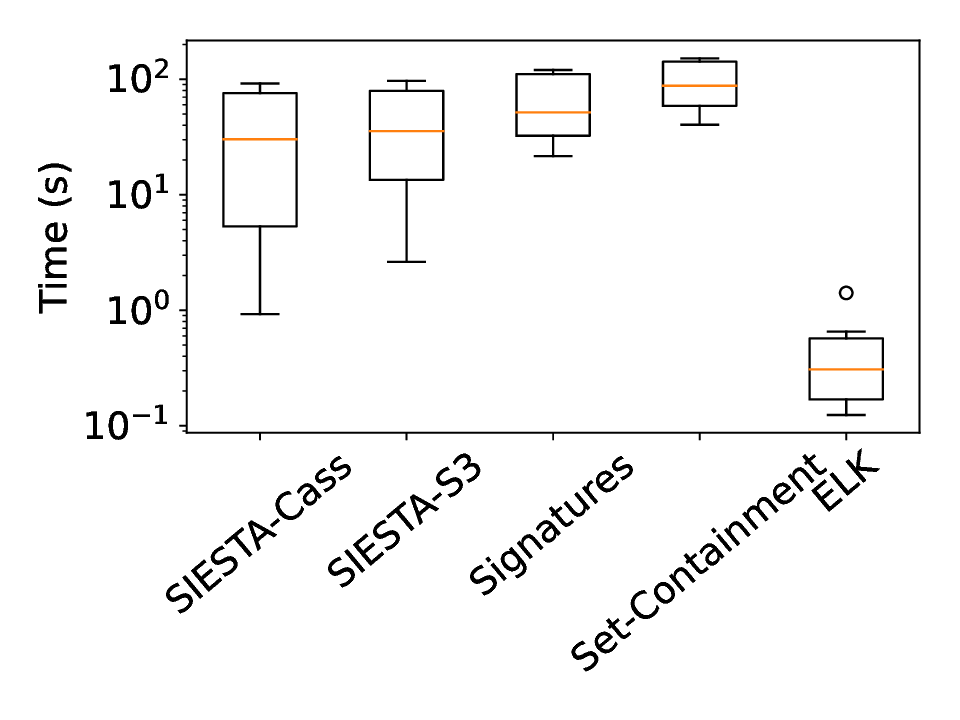}
		\caption{$|q|=2$}
	\end{subfigure}
	\begin{subfigure}[b]{0.45\columnwidth}
		\includegraphics[width=\textwidth]{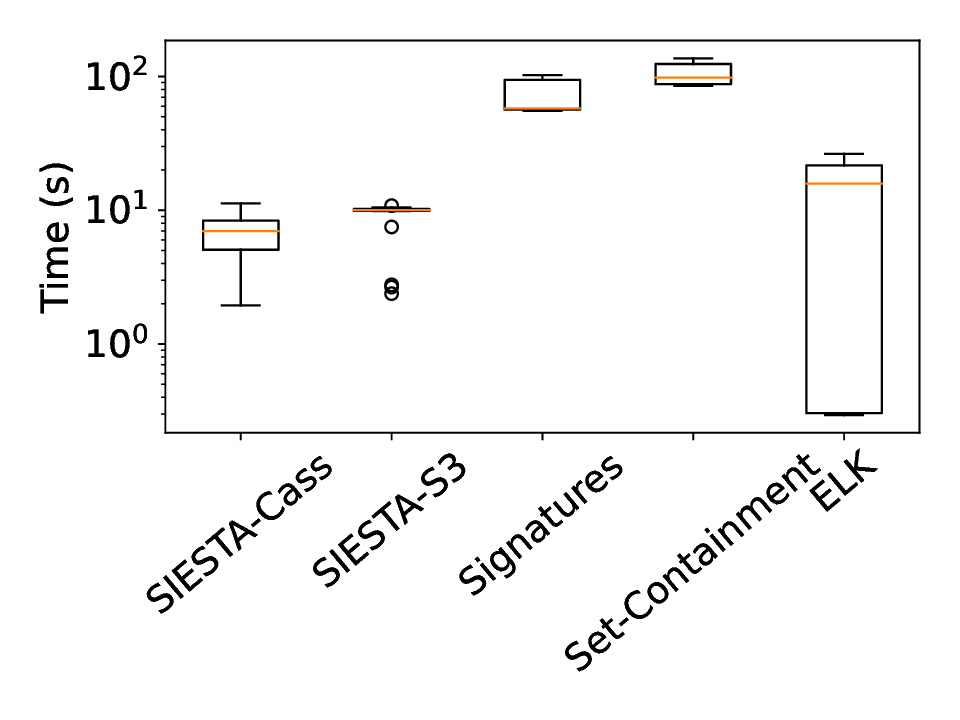}
		\caption{$|q|=10$}
	\end{subfigure}
	\begin{subfigure}[b]{0.45\columnwidth}
		\includegraphics[width=\textwidth]{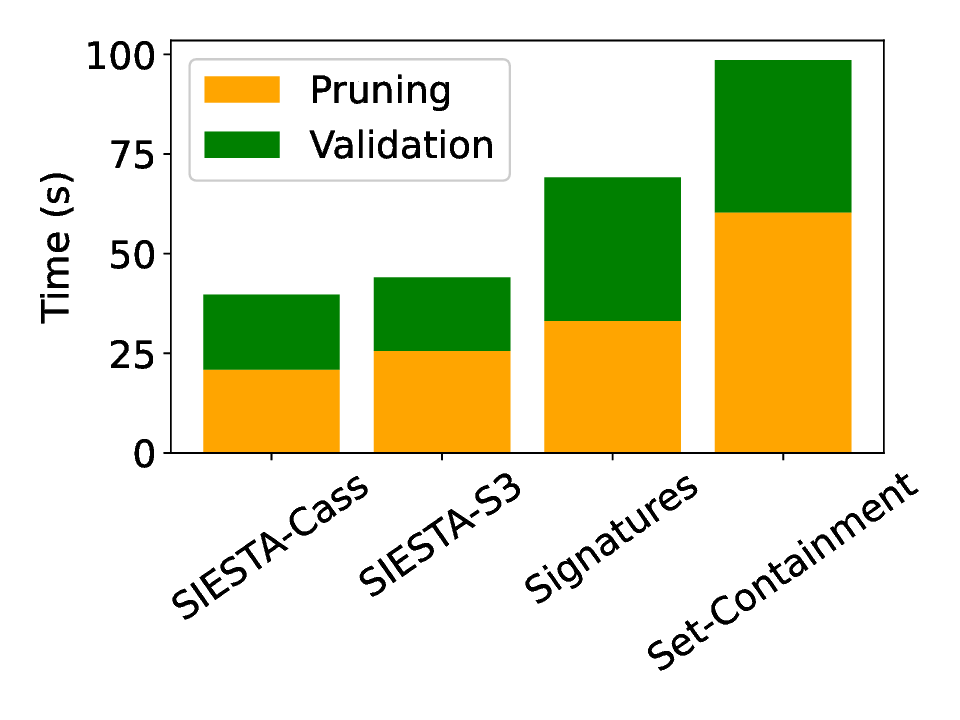}
		\caption{$|q|=2$}
	\end{subfigure}
	\begin{subfigure}[b]{0.45\columnwidth}
		\includegraphics[width=\textwidth]{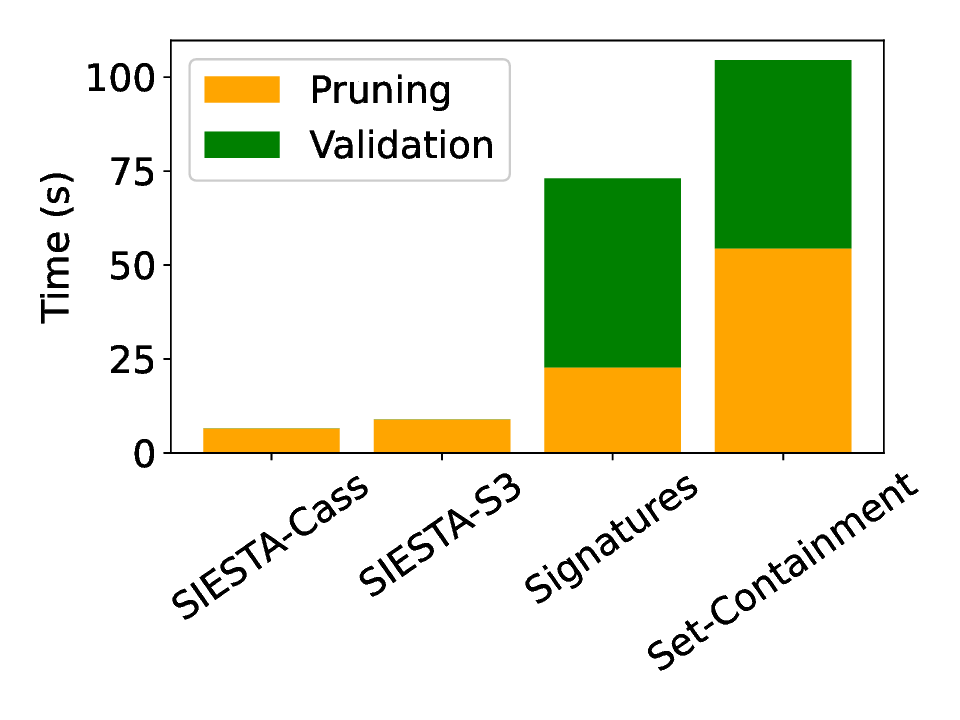}
		\caption{$|q|=10$}
	\end{subfigure}	
	\caption{Response times for simple pattern detection queries utilizing the BPI\_2019 dataset. (a,b): performance of all systems. (c,d): difference between the pruning and the validation phase of all systems implemented on top SIESTA.}
	\label{fig:q_bpi19}
\end{figure*}

Next, we evaluate the query response times for SIESTA  against the index-based competitors. Simple pattern detection queries of length 2 and 10 were executed on the BPI\_2019 dataset, without constraints or more complex operators. Figure~\ref{fig:q_bpi19} shows that ELK yields the best performance in the smaller queries, outperforming the other systems by two orders of magnitude. However, for queries of length 10, it shows a big deviation in response times with a median performance higher than SIESTA, which is rather insensitive to the event types in the query pattern. Set-Containment has the highest response times, followed by Signatures. Using Cassandra shows a greater deviation in response times, with the average time slightly better than S3. This may be due to Cassandra's caching mechanisms, allowing faster responses when data are already cached in main memory. Figures \ref{fig:q_bpi19}c and \ref{fig:q_bpi19}d further demonstrate the effective pruning of search space by SIESTA.

\begin{figure*}[tb!]
	\centering
	\begin{subfigure}[b]{0.45\columnwidth}
	\includegraphics[width=\textwidth]{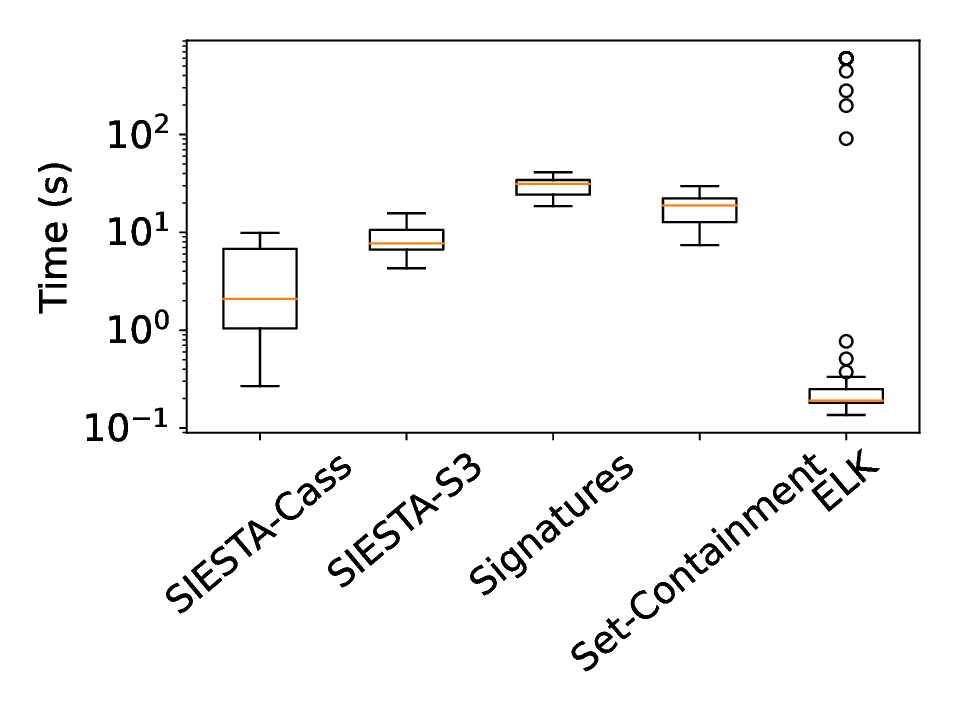}
	\caption{$|q|=2$}
	\end{subfigure}
	\begin{subfigure}[b]{0.45\columnwidth}
		\includegraphics[width=\textwidth]{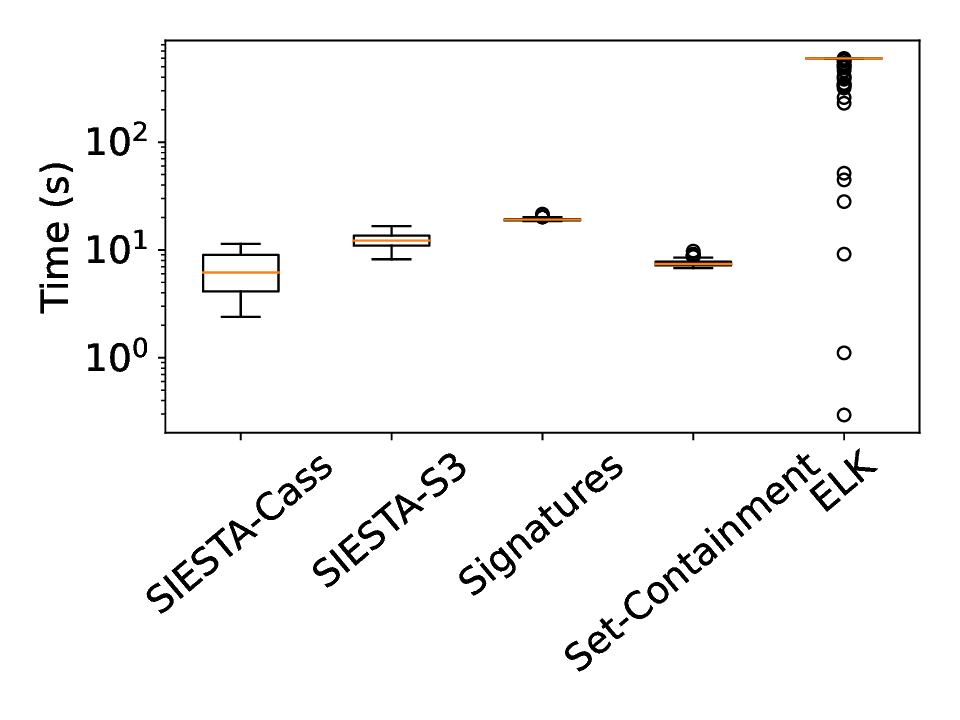}
		\caption{ $|q|=10$}
	\end{subfigure}
	\begin{subfigure}[b]{0.45\columnwidth}
		\includegraphics[width=\textwidth]{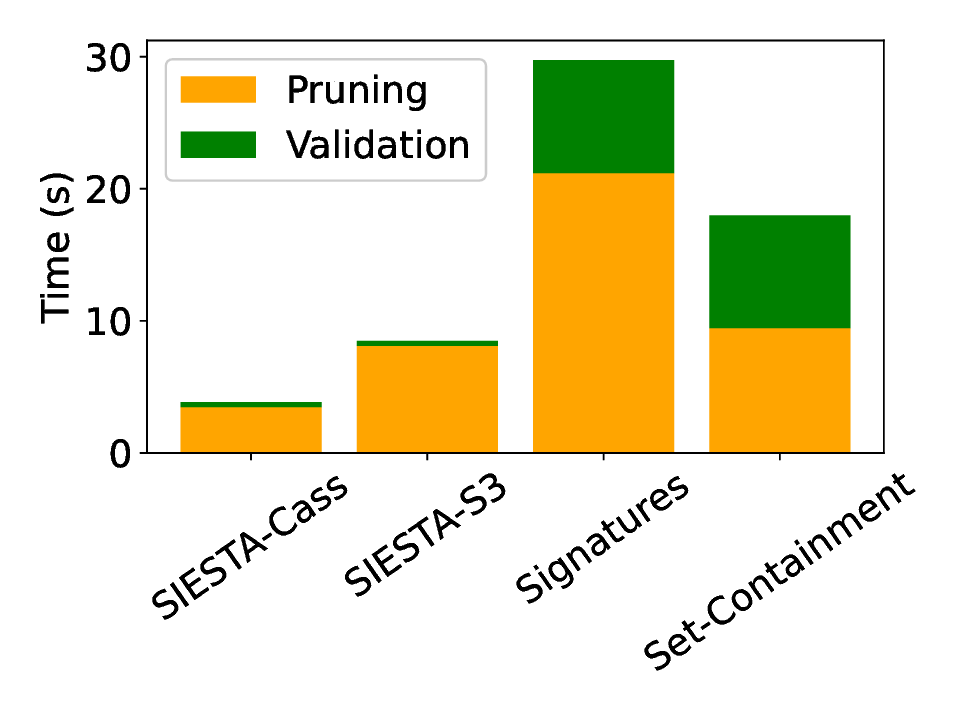}
		\caption{$|q|=2$}
	\end{subfigure}
	\begin{subfigure}[b]{0.45\columnwidth}
		\includegraphics[width=\textwidth]{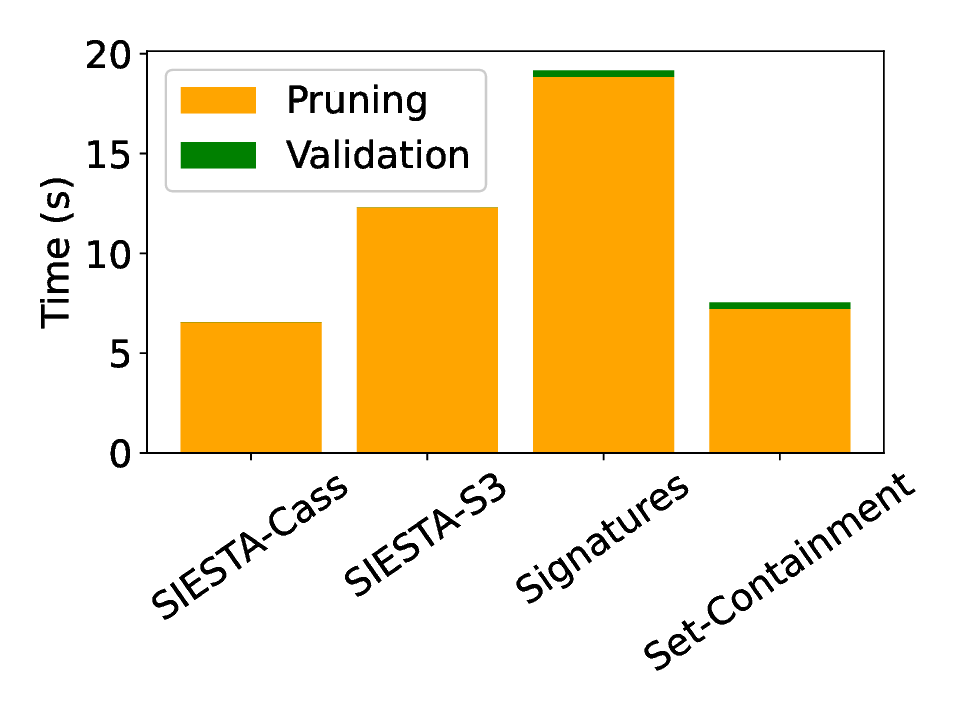}
		\caption{$|q|=10$}
	\end{subfigure}
		\caption{Response times for more complex pattern detection queries utilizing the synthetic dataset. }
	\label{fig:q_synthetic}
\end{figure*}

Following, we evaluate the performance of the various systems in more complex queries, using the synthetic dataset to see if the data distribution affects response times. While all the SIESTA-based systems can support the queries described in Section \ref{subsec:sase_integration} due to their integration with SASE, ELK does not support certain operators. According to its documentation\footnote{\url{https://www.elastic.co/guide/en/elasticsearch/reference/current/eql.html}}, ELK uses a state machine similar to CEP to detect matches. However, it cannot detect all the potential matches as this operation deemed too slow and costly to be implemented in large datasets. ELK, except for the traditional sequence operator, can also support the following operators: \emph{maxspan} (time-lapse between first and last event), \emph{until} (similar to negation, but the search stops once this event is found, while SASE will simply drop all the partial matches and continue searching) and \emph{with runs} (bounded Kleene+, between 1 and 100).
Therefore, we generated the more complex queries by randomly modifying 100 simple queries with equal parts of 15\% negations, Kleene+, and time constraints between the first and last event. Note that a query can have more than one of these operators. The results are presented in Figure~\ref{fig:q_synthetic}. 

For the smaller queries, we observe a similar behavior as before, with ELK exhibiting the best median performance. However, for bounded Kleene+ queries, ELK's average response time is 484.2 seconds, which are the outliers spotted in Figure \ref{fig:q_synthetic}a. Furthermore, for the $|q|$=10 queries, most of the ELK queries timed out at 600 seconds. This is not the case for the SIESTA-based systems that show a relatively constant behavior despite the complexity or the length of the query. Figures \ref{fig:q_synthetic}c and \ref{fig:q_synthetic}d demonstrate that Set-Containment performs well for uniformly distributed event types since the inverted lists have roughly the same size and are relatively small.

Based on the experiments described above, we can make the following observations. (1) ELK is efficient in supporting small, simple queries. However, SIESTA-based systems with SASE integration can handle a wider range of patterns and constraints, and can scale better, making them suitable for a broader range of applications. (2) Among the SIESTA-based systems, Signatures had the poorest performance, while Set-Containment has its best performance when event types are uniformly distributed. Thus, in real-world settings, our SIESTA proposal seems the most appropriate. (3)
The experiments also showed that SIESTA's indexing and pruning mechanisms are effective in reducing the search space and enabling the application of SASE on large data sets. (4) SIESTA's response times scale well with query size, regardless of the dataset's characteristics, making it a suitable choice for big data applications. 
(5) Finally, the evaluation showed that both Cassandra and S3 are viable options for storage. Cassandra provides better indexing times and slightly faster responses, but it should only be preferred when deployed in-house. On the other hand, S3 is much more cost-effective and it should be preferred for most cloud-based deployments. 

\begin{figure*}[tb!]
	\centering
	\begin{subfigure}[b]{0.49\columnwidth}
	\includegraphics[width=.8\textwidth]{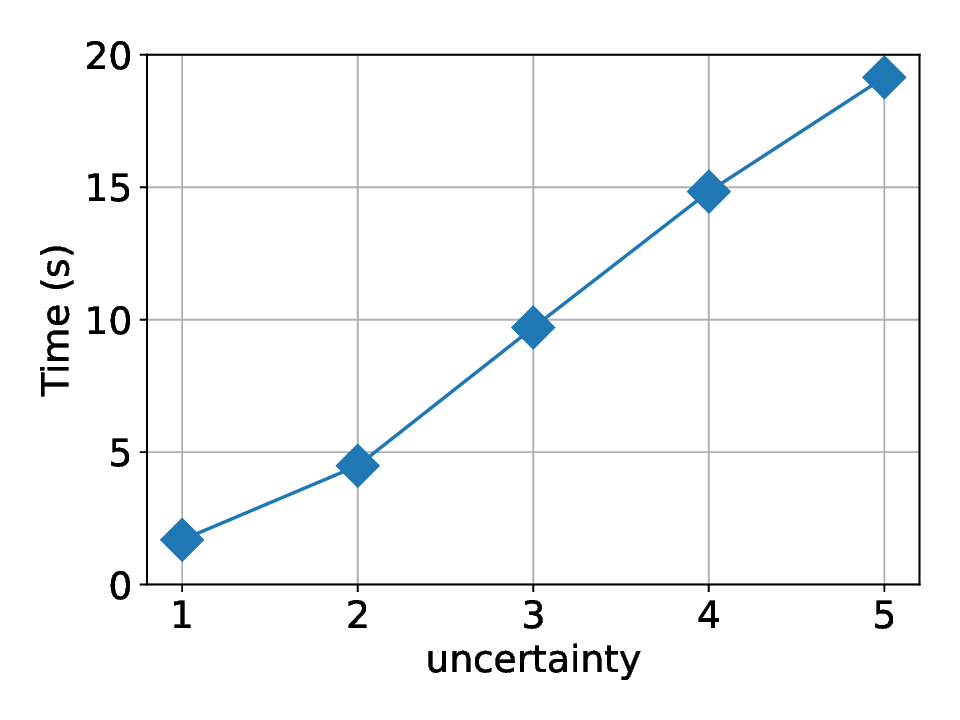}\\
        \includegraphics[width=.8\textwidth]{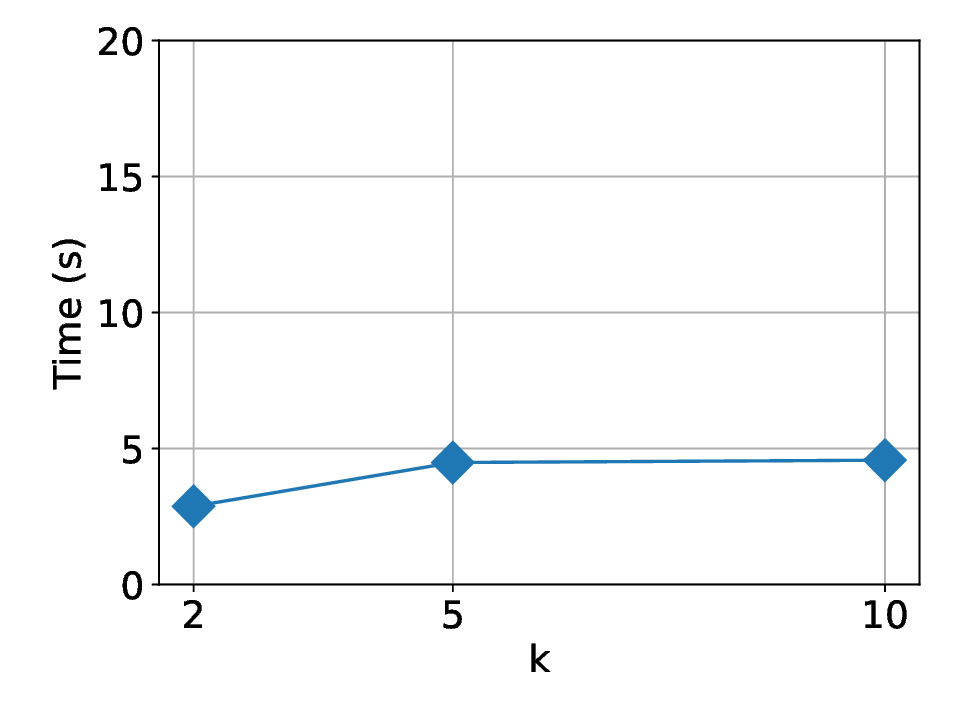}
	\caption{$|q|=3$}
	\end{subfigure}
	\begin{subfigure}[b]{0.49\columnwidth}
		\includegraphics[width=.8\textwidth]{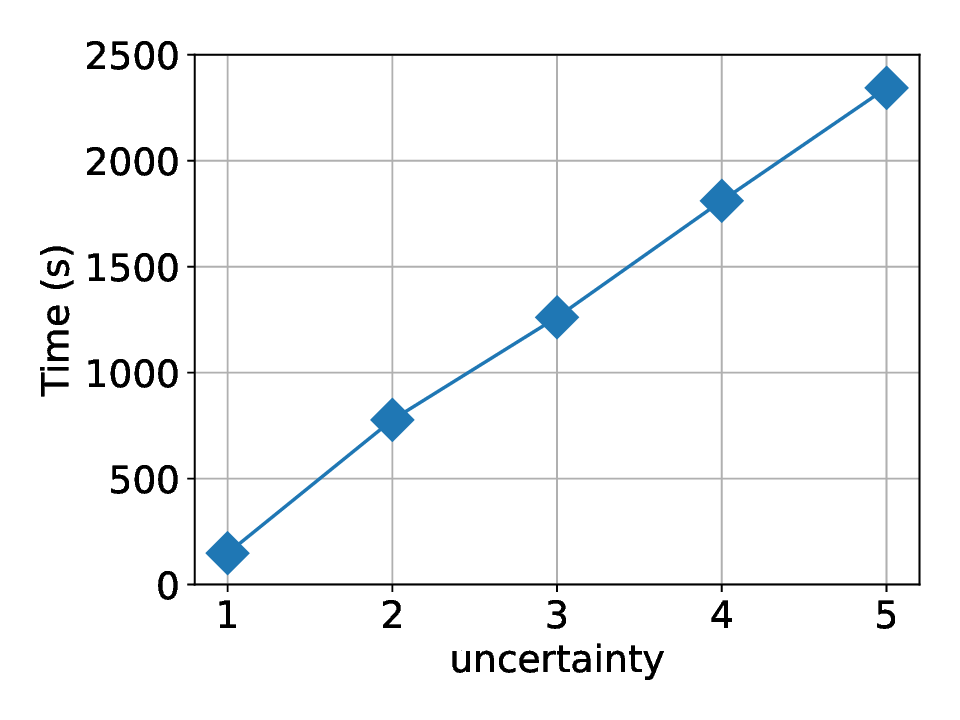}
            \includegraphics[width=.8\textwidth]{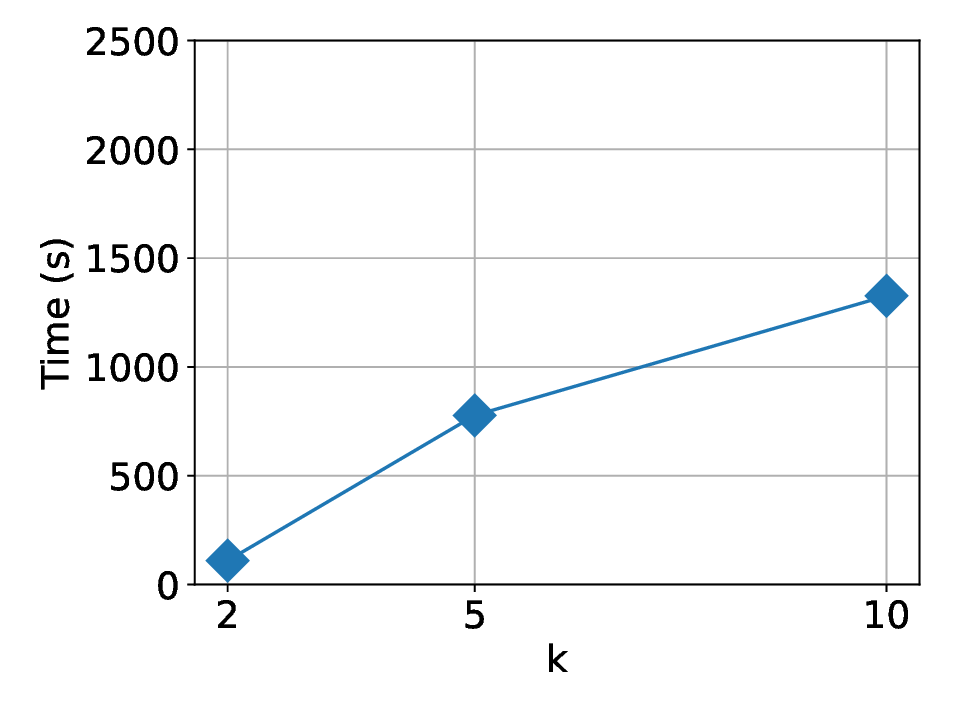}
		\caption{$|q|=5$}
	\end{subfigure}
		\caption{Response times for various parameterizations in explainability for $|q|=3$ (left) and $|q|=5$ (right). Both parameters $uncertainty$ and $k$ were set to 2 while the other was modified.}
	\label{fig:explainability}
\end{figure*}

Finally, we conducted an evaluation of SIESTA's explainability performance by executing queries of lengths 3 and 5, adjusting both $k$ and $uncertainty$. The results are presented in Figure~\ref{fig:explainability}. Notably, in both pattern lengths, the response time shows a direct correlation with the $uncertainty$ parameter. However, the $k$ parameter plays a more pivotal role in the larger queries. This is due to the increase of the possible modifications in a trace as the number of relevant events grows, and the $k$ parameter aids in the early termination of the process.
Moreover, beyond $k$ and $uncertainty$, response times are influenced by the query's length and the frequency of event occurrences within the database. As the number of events in a query increases, more events are retrieved, amplifying the impact of the $uncertainty$ parameter. Additionally, if a specific event type is infrequent in the indexed traces, even a substantial increase in $uncertainty$ has minimal impact on response times.
While explainability enhances the user experience and fosters a better understanding of query responses, it is crucial to note that it can significantly increase response times, as evidenced in the query response times for $|q|=5$. Therefore, fine-tuning is necessary to balance these trade-offs.

\section{Future Work}
\label{sec:future-work}
This work has demonstrated the ease of extending SIESTA with additional functionalities and exploring alternative implementations. Therefore, in the future, it would be interesting to further enhance the expressiveness of the supported queries, e.g. including nested operations or support for more complex data schemas with embedded domain-specific information. Additionally, exploring alternative storage options such as ScyllaDB~\footnote{\url{https://www.scylladb.com/}} and TileDB~\footnote{\url{https://tiledb.com/}} could contribute to further improving overall performance.

An intriguing idea is to combine Object Storage System (OSS) with Delta Lake\footnote{\url{https://delta.io/}} technology. This approach offers several advantages. Firstly, it enhances the performance of the incremental indexing in OSS, as it allows new events to be appended in parquet files and enforces ACID properties~\cite{armbrust2020delta}. Secondly, it allows for the seamless integration of streaming data, facilitating the combination of real-time monitoring with SIESTA's powerful pattern detection capabilities. By incorporating streaming data into SIESTA, additional analytics can be utilized such as predictive maintenance and complex event forecasting, transforming SIESTA into a complete and comprehensive system for efficiently managing logging in real-world organizations.


\section{Conclusion}
\label{sec:conclusion}
In this work, we present the SIESTA  comprehensive and scalable framework that supports cloud-native pattern detection with enhanced expressiveness, which is built on top of a preliminary proof-of-concept of work on how to index and process event logs. By integrating a state-of-the-art CEP engine, SASE, we enhance the expressiveness of SIESTA and enable the provision of explanations for unexpected results. Additionally, we explore an alternative implementation that utilizes an OSS instead of Cassandra. 
Furthermore, we provide a detailed description of SIESTA's incremental indexing approach, addressing the associated challenges. We conduct thorough experiments using a large volume of data to demonstrate the superior scalability and pattern detection capabilities of SIESTA compared to its competitors, including the state-of-the-art ELK stack and FlinkCEP. 

\emph{Acknowledgment.} 
This research was financially supported by the European Union's Horizon Europe programme under the grant agreement No 101058174 (TRINEFLEX - Transformation of energy intensive process industries through integration of energy, process, and feedstock flexibility).

\end{document}